\documentclass[review]{elsarticle}

\usepackage{bm}
\usepackage{amsmath, wasysym}
\usepackage{amsfonts}%
\usepackage{amssymb}%
\usepackage{color}
\usepackage{hyperref}
\usepackage{verbatim} 
\usepackage{epsfig,epsf}
\usepackage{graphicx,tabularx}
\newcolumntype{Y}{>{\centering\arraybackslash}X}
\usepackage{epstopdf}
\usepackage{lineno}
\usepackage{mathrsfs}
\usepackage{ulem}
\modulolinenumbers[5]

\journal{Journal of Molecular Liquids}

\begin{document}

\begin{frontmatter}

\title{Hydrogen bond correlated percolation in a supercooled water  monolayer as a hallmark of the critical region.}
%\date{\today}

\author{Valentino Bianco}
\ead{vabianco@ucm.es}
\address{Departamento de Quimica Fisica, Facultad de Ciencias Quimicas, Universidad Complutense de Madrid, 28040 Madrid, Spain}
%$^{1,2}$, 
\author{Giancarlo Franzese}
%$^1$}
\ead{gfranzese@ub.edu}
\address{Secci\'o de
 F\'isica Estad\'istica i Interdisciplin\`aria--Departament de F\'isica
 de la Mat\`eria Condensada, Facultat de F\'isica \& Institute of Nanoscience and Nanotechnology (IN2UB),
  Universitat de Barcelona, Mart\'i i Franqu\`es 1, 08028 Barcelona, Spain}
%\affiliation{$^1$Universitat de Barcelona, Diagonal 645, 08028 Barcelona, Spain}
%\affiliation{$^2$University of Vienna, Sensengasse 8/10, 1090, Vienna, Austria}
%\email{valentino.bianco@univie.ac.at}

\begin{abstract}
Numerical simulations for a number of water models have supported the possibility of a metastable liquid-liquid critical point (LLCP) in the deep supercooled region.
Here we consider a theoretical model for a supercooled liquid water monolayer   and  its mathematical mapping onto a percolation problem. The mapping allows us to identify the 
%\sout{percolating clusters} 
 finite-size clusters at any state-point, and the {\it infinite} cluster 
at the critical point,
with the regions of
correlated hydrogen bonds (HBs).
%\sout{at the critical isobar  and to discuss the relation between the percolation critical exponents and their thermodynamic  corresponding exponents.} 
We show that the percolation line coincides with the first-order liquid-liquid phase transition ending at the LLCP.  
 At pressures below the LLCP, the percolation line corresponds to the strong maxima of the thermodynamic response functions and to the locus of maximum correlation length (Widom line).
At higher pressures, we find a 
%\sout{random-bond}
percolation transition with a positive slope and we discuss its possible relation with the thermodynamics.
%\sout{Our approach allows us to show that the correlation length of the correlated HBs  diverges approaching the LLCP in the thermodynamic limit, a result that is difficult to prove by direct calculation in water models.}
\end{abstract}

%\maketitle


\end{frontmatter}

\section{Introduction}

A fundamental feature of water molecules is their capability to form hydrogen bonds (HBs) \cite{Ball:2014aa}. In the liquid phase the HBs organize in a quasi-tetrahedral network that is continuously built up and broken on timescales of picoseconds \cite{Fernandez-Serra2006}. Although there is scientific consensus about the relevance of the HBs for the anomalous properties of water, the consequences of the HB peculiar properties on the supercooled water phase diagram are under 
 debate  since decades \cite{Kanno:1975aa, Kanno1979, Mishima1985, llcp, Mishima1998, Xu2005, FS2007, Angell2008, Mallamace2008, Kim:2009aa, Stokely2010, Mazza2011, Nilsson:2012fk, limmer2013, Gallo2013, Palmer:2014uq, SmallenburgPRL2015, Smallenburg:2015aa,  DeMarzio2016, Gallo:2016aa, Menzl2016, Rovigatti2017, Perakis:2017aa, Handle2017, Russo2018, Ni:2018aa, Palmer:2018aa, Palmer:2018ab}.
%\sout{In bulk water near freezing each molecule minimize the free energy by favoring a hydration shell made of only four H-bonded molecules, while at higher temperature the free energy is dominated by the entropy that increases by increasing the local density, with five of six nearest neighbors and a decreasing number of HBs.}
 In bulk water near freezing each molecule minimize the free energy by favoring a hydration shell made of four tetrahedral H-bonded molecules. 
For decreasing temperature $T$, this mechanism leads to increasing fluctuations in volume $\langle V^2\rangle$ and in energy $\langle E^2\rangle$--associated to the isothermal compressibility $K_T$ and the  isobaric specific heat $C_P$, respectively-- and to increasing cross-fluctuations in volume and energy, related to the isobaric thermal expansion coefficient $\alpha_P$.
The thermodynamics response functions $K_T$ and  $C_P$ increase  below 46 and 35 $^o$C, respectively,  while $\alpha_P$ becomes negative below 4 $^o$C and increases in absolute value  \cite{Kell1975, Angell1973, SPEEDY1976, AngellC.A._j100395a032}. However, their largest increase occur 
 in the supercooled region, below the melting line, where water can be liquid down to temperatures between -46 and -42 $^o$C, approximately \cite{Sellberg:2014fk, Kim:2017aa, PhysRevLett.120.015501}. Similar results have been observed at negative pressure and moderate supercooling \cite{Holten:2017aa}.

The rapid increase of the fluctuations  has suggested the possible existence of a divergency at approximately -45 $^o$C \cite{SPEEDY1976}, below or near the  experimental limit at which the present technology allows us to probe the metastable liquid \cite{Sellberg:2014fk, Kim:2017aa, PhysRevLett.120.015501}. 
This possibility has been supported by the seminal work  of Poole et al. in 1992, in which the  existence of a liquid-liquid critical point (LLCP) in the deep supercooled regime was suggested based on numerical simulations of the ST2 model \cite{llcp}, one of the many atomistic models for water. 
The LLCP hypothesis  since then has been corroborated by a series of theoretical and experimental results \cite{Fuentevilla2006, Abascal2010, Stokely2010, Kesselring2012, Liu2012, Holten2012b, Bianco2014, Sciortino2011, Kesselring2013, Palmer:2014uq, SmallenburgPRL2015, Ni2016, Palmer:2016ab, Handle2017}, although different scenarios are still under debate \cite{Speedy:1982aa, 
Sastry:1996aa, Angell2008, Stokely2010, Limmer2011}. One of them \cite{Limmer2011, limmer2013} recently was invalidated due to conceptual errors at its origin \cite{Palmer:2018aa, Palmer:2018ab}, while another \cite{Speedy:1982aa} was found in a colloidal model \cite{Rovigatti2017}, but does not seem to be compatible with recent experimental data for water \cite{Kim:2017aa, Holten:2017aa}.

The LLCP scenario hypothesizes the existence of a first order phase transition line separating two metastable liquid phases, the high-density liquid (HDL) at higher pressures and temperatures, and the low-density liquid (LDL) at lower pressures and temperatures. 
The presence of a critical point (CP) would imply the existence of the Widom line  \cite{Xu2005, FS2007, Luo2014, Holten2012a}, that emanates from the CP and is the locus in  the $T$--$P$ plane  where the statistical correlation length $\xi$ has a  maximum.  The Widom line is calculated along the constant thermal field \cite{Luo2014, Holten2012a}  and not necessarily coincides with the maxima of $K_T$, $C_P$ or $\alpha_P$ \cite{doi:10.1021/jp2039898}, neither represents a unique thermodynamic separatrix of the supercritical region \cite{PhysRevE.98.022104}. A detailed definition of the Widom line is presented in the Appendix.
The  $\xi$ maximum along the Widom line increases approaching the CP and diverges at the CP.

Nonetheless, no experiment so far have found a direct  evidence of the LLCP in supercooled water. The reason is that the LLCP, if present in bulk water, should occur where the lifetime of the metastable liquid is too short for being probed with the present experimental techniques. Therefore, different strategies have been developed to perform experiments that could help us in establishing if the LLCP occurs in bulk water. Among them water solutions \cite{Woutersen:2018aa} or mixtures \cite{Murata:2012fk} have been used, as well as water under confinement \cite{Bertrand2013}.

However, confinement-dependent effects on the LLCP are difficult to predict without detailed studies.  In the following we study the case of 
 a supercooled single layer of water confined in two-dimensional geometry. 
To this goal, we investigate a coarse-grain water model by means of a percolation mapping that allow us to identify the critical region of the system. 

Such a description is based on a mathematical mapping developed by Kasteleyn and Fortuin \cite{Kasteleyn_Fortuin}, and Coniglio and Klein \cite{coniglio1979}--namely the site-bond correlated percolation--between physical and geometric properties \cite{Kasteleyn_Fortuin, Fortuin1972536, coniglio1979, coniglioJPhysA1980, Cataudella1996, Franzese1996, Franzese1998, Fierro1999, FC1999, Franzese2000} which guarantees that the correlation length $\xi$ statistically coincides with the connectivity length $\xi_c$, i.e. the average size of the cluster.
 
A few works studying the percolation region in the super-critical liquid water have been published so far \cite{partay:024502, B617042K, bernabeiPRE2008, Strong:2018aa}. They are  based on a definition of a cluster as  a contiguous region of H-bonded molecules. As we will describe in the next sections, this definition provides an overestimation of the region of statistically correlated molecules and cannot be stricly related to the thermodynamic behavior of water.  Indeed, the $T$-$P$ locus of the percolation transition and its nature change according to the definition of cluster of molecules, which is in principle arbitrary. For example, it is possible to define a cluster as the contiguous region of molecules whose distance is below a given threshold, or as the contiguous region of molecules bonded via HB, etc. Each definition is rational and identifies a specific feature of the system, but none of them is in principle related to the size $\xi$ over which the statistical fluctuations of the degrees of freedom spread.
On the other hand, the Kasteleyn-Fortuin/Coniglio-Klein approach (described in the section 3.2), allows to define a probability $p$ that two degrees of freedom belong to the same cluster, in such a way that the critical thermodynamic behavior is preserved in the percolation description or, in other words, that the clusters have a characteristic length-scale (connectivity length) that coincides with $\xi$.

Here, hence, we characterize the formation of the HB network in relation with the clusters of correlated water molecules. According to the $T$ and $P$ conditions, we compute the mean cluster-size and cluster-size distribution, identifying the critical region where the clusters span all over the system and mark the onset of a percolation transition. We show how the percolation transition is associated to the building up of the ordered HB network, although with different mechanisms at higher and lower pressures. 
%\sout{Moreover, we show how the mean cluster-size captures the anomalous increase upon cooling of the isothermal compressibility and of the isobaric specific heat upon cooling.}
Finally, we discuss the low-$P$ percolation transition in relation to the Widom line in supercooled 
water.
%\sout{or to a possible critical transition induced by the confinement.}

 \section{The FS model for a water monolayer} 
 
We adopt   the
%\sout{a} 
{\it many-body}
model for a water monolayer %\sout{that has been}
introduced in 2002 by Franzese and Stanley 
 (FS)
\cite{Franzese:2002aa, FS-PhysA2002} 
\sout{and} that has been shown to reproduce, at least qualitatively, many of the water properties
\cite{FMS2003, Stokely2010, KFS2008, delosSantos2011, delosSantos2012, Mazza2009, strekalovaPRL2011, Franzese2011, Bianco2012a, Mazza2011, Mazza2012, Bianco2014, Bianco:2015aa, Bianco:2017ab, BiancoPRX2017}. 
 When the many-body interaction is zero the FS model coincides with the model introduced in Ref.\cite{Sastry:1996aa} by Sastry, Debenedetti, Sciortino and Stanley.
%\sout{Simulations with atomistic water models show that water crystallization is inhibited, e.g., when water is confined  between parallel plates of quartz at a distance less then 0.5 nm, possibly related to experimental results at $T\simeq -8^o$C \cite{Zangi2003}, or when water is confined between parallel sheets of graphene at a distance less then 0.7 nm near 0$^o$C \cite{MCF2017}. In the present model we assume that the confinement always keeps the water monolayer liquid. As we will discuss in the following, such assumption is drastic, but allows us to simplify the model and to explore by theoretical methods the thermodynamic behaviour of the supercooled liquid at extreme conditions before it would crystallize.}
The monolayer is kept between parallel walls at sub-nanometer distance, $h=0.5$ nm,  with purely repulsive (excluded volume) hydrophobic interaction,
allowing deep supercooling without crystallization, consistent with simulations and experiments \cite{Zangi2003, MCF2017}.

We consider 
%\sout{a monolayer at constant temperature $T$, pressure $P$ and number of molecules $N$, while the volume $V_{\rm tot}$ has a fixed hight $h=0.5$ nm and can fluctuate in the two directions parallel to the confining walls. The total volume changes with $T$ and $P$ and  can be decomposed as $V_{\rm tot}=V+V_{\rm HB}$, where $V$ is independent of the total number $N_{\rm HB}$ of HBs made by the water, and a HB-dependent contribution $V_{\rm HB}$. The interaction with the walls is purely repulsive (excluded volume) as in an idealized hydrophobic surface.}
 the model in the Gibbs ensemble, keeping the number $N$ of molecules, the pressure $P$ and the temperature $T$ fixed. Therefore, the volume $V$ changes according to the equation of state and fluctuates when we perform our Monte Carlo simulations. Assuming that  the density is homogenous, this implies that the (average) distance $r$ between the molecules changes with $P$ and $T$. We partition the system into cells whose volume coincides with the (average) volume occupied by each molecules. Hence, by construction, each cell has the size equal to the inter-molecule distance $r$ and the cell's size changes with $P$ and $T$.
Because we consider here a monolayer, we adopt a square partition, whose coordination number  coincides with the number of HBs that each molecule forms at low temperature. In three dimensions 
other partitions are more appropriate \cite{2016arXiv161000419C}.

To reduce the number of degrees of freedom of the system,  in accordance with FS, we coarse grain
the molecules translational coordinates 
%
%\sout{ are coarse-grained by a density field with a resolution given by cells with hight $h$ and square surface of size $r^2\equiv v/h \equiv V/h/N$, with $r$ corresponding to the average  O--O distance between the molecules in absence of HBs. The single cell volume $v=V/N$, independent of $N_{\rm HB}$, is  $v\geq v_0\equiv h r_0^2$, where  $r_0\simeq 2.9\AA $  is the van der Waals diameter of a water molecule. We consider here the homogenous case in which each cell is occupied by a water molecules and we associate a discrete variable $n_{i}=1$ if the cell $i$ has $v_0/v>0.5$, and $n_{i}=0$ otherwise.  The case $n_{i}=0$, hence, corresponds to $v\geq 2 v_0$ and $r\geq \sqrt{2} ~2.9 \AA \simeq 4 \AA\equiv r_{\rm max}$. This quantity  is in fair agreement with the maximum elongation, $\simeq 3.7 \AA $,}
%
  as follows. 
By construction each cell includes always a water molecule and its volume $v$ is at least equal to the van der Waals volume $v_0$ of the molecule. Hence, $v_0/v$ is the cell density in van der Waals units. We associate an index $n_i=0$ if $v_0/v\leq 0.5$. For our choice of geometry and parameters ($v_0\equiv hr_0^2$ with $r_0\simeq 2.9\AA $ van der Waals diameter), it is  $n_i=0$ when 
$r\geq \sqrt{2} ~2.9 \AA \geq 3.7 \AA  \equiv r_{\rm max}$ 
 maximum O--O elongation 
of a straight HB as calculated by {\it ab initio} molecular dynamics simulations from the proton-transfer coordinate \cite{Ceriotti:2013aa}, assuming  a covalent distance O--H $\simeq 1 \AA $.
Therefore,  a cell with $n_i=0$ cannot form HBs, although it includes a molecule. On the other hand, we associate to the cell an index $n_i=1$ if $v_0/v> 0.5$, i.e. $r\leq r_{\rm max}$, and the molecule in it can form HBs. Hence, the index $n_i$ is a discretized density field that marks if the molecule $i$ can, or cannot, form HBs. If we consider that all the HBs are formed in the liquid phase, then $n_i=1$ is associated to a liquid-like density, and $n_i=0$ to a gas-like density, recalling a lattice-gas model for argon-like systems. Hence,  in the liquid phase, all the indices $n_i$ are equal to 1, while they are all 0 in the gas phase. Therefore, the average $\langle n_i \rangle=n_i$  does not play the role of an order parameter as in the lattice-gas model. Instead, here the order parameter for the liquid-gas transition is associated to the total density, that, as the total volume, is a continuous function of $P$ and $T$.

Furthermore, water unlike argon and other usual liquids, has HBs. Following FS,
we assume that all the possible heterogeneities of water are due to the HBs.
When water forms HBs within its hydration shell, its coordination number reduces to 4, in an almost perfectly tetrahedral configuration. In this fully-bonded configuration the volume per molecule is larger than in configurations with larger coordination number, because the volume occupied by the hydration shell as a whole is not changing in a sensible way, while the number of molecules in it does change \cite{Soper-Ricci-2000}. This can be represented, approximately, by associating a {\it proper} volume $v_{\rm HB}$ to each new HB -- the larger the number of HBs in the system, the better the approximation. 
We choose $v_{\rm HB}/v_0=0.5$, equal to the average volume increase between high-$\rho$ ices VI and VIII and low-$\rho$ (tetrahedral) ice Ih.

As a consequence, since the number $N$ of molecules is fixed, the total volume $V_{\rm tot}$ occupied by the system increases linearly with the number of HBs, $N_{\rm HB}$, i.e., 
\begin{equation}
V_{\rm tot}\equiv V+N_{\rm HB}v_{\rm HB}, 
\end{equation}
where $V\equiv Nv$ is the volume without HBs. Despite $V$ is homogeneously distributed among the $N$ water molecules, the number of HBs changes among molecules, giving rise to local heterogeneities in the density field.

%Moreover, we associate 4 bonding variables to each molecule to describe the HB interaction and its effect on the local volume, as described in the following. In this way, the Cartesian coordinates of the molecules are replaced by the sum of a continuous density field and discrete variables.

The Hamiltonian of the   FS model is given by 
\begin{equation}\label{hamiltoniana}
 \mathscr{H}\equiv \sum_{ij} U(r_{ij}) -JN_{\rm HB} - J_{\sigma}N_{\rm coop}.
\end{equation} 

The first term, summed over all the water molecules $i$ and $j$ at O--O distance $r_{ij}$, is given by 
 a truncated Lennard-Jones (LJ) model
\begin{equation}
 U(r)\equiv
 \begin{cases}
\infty\text{ for } r<r_0\\\\
4\epsilon \left[\left(\dfrac{r_0}{r}\right)^{12}-\left(\dfrac{r_0}{r}\right)^6\right]  \text{ for }  r_0 \leq r\leq 25r_0\\\\
0 \text{ for } r > 25r_0
 \end{cases}
 ,
\end{equation} 
where $\epsilon \equiv 5.8$ kJ/mol, close to the estimate  based on isoelectronic molecules at optimal separation  $\simeq 5.5$ kJ/mol \cite{Henry2002}. The potential 
$U(r)$ takes into account the van der Walls (dispersive) attraction and hard core (electron) repulsion between water molecules. 
%\sout{It represents the isotropic component of the water-water interaction}

As already noted, the formation of HBs, does not affect the volume occupied by the full-bonded hydration shell, but only the number of molecules included in it. Therefore, the distance $r$ between the molecules is not modified by the HBs \cite{Soper-Ricci-2000}. Hence, the van der Waals interaction is not affected by the HBs \footnote{We acknowledge discussion with the late Prof. David Chandler for noticing that this point is crucial to state that the FS model is not mean field.}. Because in the FS model $r$ is a continuous variable, it is appropriate to represent the van der Waals interaction with a  LJ model. We truncate the LJ potential at large distances, as usually done in continuous models for numerical efficiency, and also at short distance. In particular, we replace the repulsive power 12 \footnote{While the power 6 of the LJ model can be derived from first principke calculations, the power 12 is arbitrary.} with a hard-core at the van der Waals diameter. Our previous analysis show that both truncations do not affect the results and simplify the implementation of the model
\cite{Santos:2011aa}. The truncated LJ potential drives the liquid-gas phase transition. It plays a  fundamental role in the vicinity of the liquid-gas spinodal and is the only relevant interaction for temperatures above the spinodal temperature.
Without this term the model would not reproduce correctly the fluid phases of water.

%{\color{blue} and is responsible for the liquid-gas transition in the model. It is important to stress that, even though our coarsening consist in discretizing the volume in cells, this do not imply that the cell size $r$ is fixed. In particular, when the system is in the liquid state, all the $n_i=1$, and the cell partitioning impose only that, for any size $r$, we have a fixed number of pair molecules at a given distance in unit of $r$. In other words, for any value of the cell size $r$ we have $2N$ pairs at distance $r$, $2N$ pairs at distance $\sqrt{2}r$, $2N$ pairs at distance $2r$, $4N$ pairs at distance $\sqrt{5}r$, and so on. Along the simulation, $r$ fluctuates according to the Boltzmann weight due to the $U(r)$ term, involving a change of $V$ and $E$. }

The second term 
 in Eq.(\ref{hamiltoniana})
represents the short-range, directional, covalent \cite{Galkina:2017aa} component of the HB, with $J/4\epsilon=0.5$, i.e. $J\simeq 11$ kJ/mol, close to the estimate 
 of this energy constant
that can be derived from the optimal HB energy and a HB cluster analysis  \cite{Stokely2010}.
%\sout{, and where}
 Here
\begin{equation}
 N_{\rm HB}\equiv \sum_{\langle ij \rangle} n_i n_j \delta_{\sigma_{ij},\sigma_{ji}},
\end{equation} 
%\sout{is}
the number of HBs,
 is
%Here we introduce the label $n_i = 1$ if the cell $i$ has a water density $N/V> 0.5/v_0$ and $n_i = 0$ otherwise.
%\sout{Here}
the sum 
%\sout{is}
over nearest neighbors (n.n.) 
 pairs of water molecules $i$ and $j$ at a distance $r\leq r_{\it max}$ \cite{Soper1986, Luzar-Chandler96} (i.e., with $n_i n_j=1$) and in the same {\it bonding state} ($\delta_{\sigma_{ij},\sigma_{ji}}=1$), where 

%\sout{and}
$\sigma_{ij} = 1, \dots, q$ is the  {\it bonding variable} %\sout{index}
of molecule $i$ 
 facing
%\sout{to}
the n.n. molecule $j$, with $\delta_{ab}=1$ if $a=b$, 0 otherwise. 

 The bonding variables are introduced to account correctly for the variation of entropy and energy associated to the formation or breaking of a HB, as explained in the following.
If two n.n. molecules $i$ and $j$ form a HB, the system energy decreases by $-J$ and  the system entropy decreases by $-k_B \ln q$ ($k_B$ is the Boltzmann constant) because both molecules have  bonding variables in the same state, $\sigma_{ij}=\sigma_{ji}$.

Each molecule has 4 bonding %\sout{indices} 
 variables, one for each possible HB, and $q^4$ possible bonding  
 configurations.
%\sout{states. For each new HB, a bonding index is fixed and the molecule looses $q$ accessible states, accounting for the entropy loss due to HB formation.} 
%\sout{By choosing $q=6$ we guarantee the correct HB definition and entropy loss. Indeed,}
The HB 
between two oxygens
is broken if ${\widehat{\rm OOH}}> 30^\circ$, 
 or ${\widehat{\rm OOH}}<- 30^\circ$,
as estimated from Debye-Waller factors \cite{Teixeira1990, Luzar-Chandler96}.
%\sout{and} 
Therefore, only $1/6$ of the entire range of 
possible orientations $[0,360^\circ]$ in the OH---O plane is associated to a bonded state,
and each HB formation leads to an entropy decrease equal to $-k_B \ln 6$. Therefore, by choosing $q=6$ we guarantee the correct HB definition and entropy loss.

%\sout{, i.e., for each possible HB there is only 1 over 6 states that is bonded. Note that the factor $n_i n_j$ in this term guarantees that the HB can be formed only when the O-O distance is $< r_{\rm max}$} 
%\cite{Soper1986, Luzar-Chandler96}.
% {\color{blue} In such a way, the HB formation is coupled to the fluctuation of the cell size $r$ due to the $U(r)$ term.}

The third term in Eq.(\ref{hamiltoniana})
accounts for  the HB cooperativity due to O--O--O correlation, that in bulk leads the molecules toward an ordered tetrahedral configuration  \cite{Soper-Ricci-2000}. Such an effect originates from quantum many-body interactions of the HB \cite{Cooperativity,  Hernandez-de-la-Pena:2005vn}. The number of cooperative 
%\sout{interactions}
 pairs of bonding variables in the system is
\begin{equation}
 N_{\rm coop}\equiv \sum_i n_i\sum_{(l,k)_i}\delta_{\sigma_{ik},\sigma_{il}},
\end{equation} 
where, 
 for each molecule $i$,
$(l,k)_i$ indicates each of the six different pairs of the four indexes $\sigma_{ij}$ of the molecule.

Therefore, the enthalpy of the system can be written as 
\begin{equation}
% \begin{array}{ll}
%& 
H\equiv U(v) -J_{\rm eff}\sum_{\langle i,j \rangle}n_in_j\delta_{\sigma_{ij},\sigma_{ji}} 
%\\\\ & 
-J_\sigma \sum_i n_i\sum_{(k,l)_i}\delta_{\sigma_{ik},\sigma_{il}}+Pv,
% \end{array}
\end{equation} 
where $J_{\rm eff}\equiv J-Pv_{HB}$
is the effective interaction between $\sigma$-variables of n.n. molecules that depends on $P$ 
and 
$U(v) \equiv \sum_{i,j}U(r_{ij})$, where
$r_{ij}$ is a function of $r=\sqrt{Nv/h}$: $r_{ij}=r$ if $i$ and $j$ are n.n., $r_{ij}=\sqrt{2}r$ if $i$ and $j$ are next n.n., etc.

In the $NPT$ ensemble the partition function of the system is 
\begin{equation}\label{eq1}
 Z(T, P)\equiv \sum_{\{\sigma\}\{v\}} e^{-H/k_BT}, 
\end{equation} 
where the sum is over all the possible configurations of bonding variables $\{\sigma\}$ and cell 
 volumes $\{v\}$.
The Eq. (\ref{eq1}) can be rewritten as
\begin{equation}
  Z(T, P)= \sum_{\{v\}} e^{- [U(v) + Pv]/k_BT} \times Z_{\{\sigma\}}
\end{equation}
where\\
\begin{equation}\label{eq2_bis}
 \begin{array}{llll} 
Z_{\{\sigma\}} & \equiv & \sum_{\{\sigma\}} & e^{(J_{\rm eff}/k_BT)\sum_{\langle i,j\rangle}n_in_j\delta_{\sigma_{ij},\sigma_{ji}}}  
%\\\\ & 
~\times~ 
e^{(J_\sigma/k_BT) \sum_i n_i\sum_{(k,l)_i}\delta_{\sigma_{ik},\sigma_{il}}  }
\\\\ & = &
\sum_{\{\sigma\}} & 
\prod_{\langle i,j\rangle} \left[ 1 + \left( e^{(J_{\rm eff}/k_BT)}-1\right)n_in_j \delta_{\sigma_{ij},\sigma_{ji}} \right]  
\\\\ & ~ & ~ &
\times \prod_{i=1}^N \prod_{(k,l)_i} \left[ 1 +\left(e^{(J_\sigma/k_BT)}-1\right) n_i \delta_{\sigma_{ik},\sigma_{il}}\right]
 \end{array}
\end{equation}
where $\prod_{\langle i,j\rangle}$ runs over all the n.n. molecules $j$ of the molecule  $i$, $\prod_{i=1}^N$ runs over all molecules and $\prod_{(k,l)_i}$ extends over all the six pairs of  bonding variables of a specific molecule $i$.

%\sout{Here is}
 In the following we set
 $J_{\sigma}/4\epsilon=0.05$, i.e. $J_{\sigma}\simeq 1$ kJ/mol,
in such a way that, for each molecule whose bonding configuration changes from three bonding variables in the same state to four (adding a cooperative HB), the energy decreases by $-3J_{\sigma}\simeq -3$ kJ/mol.  
This value is consistent with the energy decrease 

%\sout{that is the estimate of the cooperative interaction among HBs if the strength increase} 
of -3 kJ/mol \cite{Eisenberg1969} of each HB observed in ice Ih with respect to liquid water, 
in the reasonable hypothesis that this energy change
is 
entirely
attributed to the HB cooperativity \cite{Heggie1996, Stokely2010}.
 Our choice of the model parameters, with
%\sout{The value}
$J_{\sigma}\ll J$, guarantees an asymmetry between the two components of the HB interaction, such that bonding %\sout{indices} 
variables
can organize cooperatively only if each single HB is formed. 
From here on we express $T$ and $P$ in internal units, $4\epsilon/k_B$ and $4\epsilon/v_0$ respectively.
%\sout{being $k_B$ is the Boltzmann constant.}

\section{Site-bond correlated percolation } 

\subsection{Percolation mapping}

The 
%\sout{general problem of  the} 
percolation theory 
studies the connectivity properties of clusters defined
%\sout{concerns the study of {\it geometrical objects} randomly placed, with an occupation probability $p$,} 
on a $d$-dimensional lattice. 
We talk of {\it site} or {\it bond} percolation depending on whether the clusters are made of n.n. vertices or links of the lattice, respectively, that have been randomly occupied with a probability $p$.  

%\sout{If two placed object are n.n., they are connected. These objects have a characteristic connectivity radius $r$ such that two objects are, on average, connected by a path of placed objects if their distance is smaller than $r$. A contiguous region of connected objects defines a cluster. The percolation theory studies the geometrical properties of such clusters.} 
For each lattice size $L$, 
%\sout{a special value of $p^*(L)$ is} 
the smallest 
value of $p$
at which 
there is 50\% probability of finding
at least a cluster 
%\sout{spans} 
spanning
the entire lattice
is indicated as $p^*$.
%\sout{, e.g., connecting two opposite lattice sides.} 
The thermodynamic limit $p^*(L\rightarrow \infty)\equiv p_c$ is called the {\it percolation threshold}.

 The theory shows
%\sout{It is well known} 
that the percolation 
%\sout{problem} 
on large lattices displays 
%\sout{the features of a system undergoing}
a second-order phase transition where $p$ is the control 
%\sout{parameter, corresponding to the temperature in a thermodynamic system, while}
field and
 the order parameter is the probability  $\mathscr{P}_\infty$ for an arbitrary lattice site to belong to the percolating cluster \cite{stauffer1994introduction, Coniglio2016}. 
%\sout{Indeed, it is possible to show that exists an exact mapping  define percolation quantities which diverge (or vanish), with characteristic} 
The percolation transition is characterized by
critical exponents 
for $p\rightarrow p_c$, in analogy with  thermodynamic second-order phase transitions where $p$ is replaced by the thermal field \cite{StanleyBook,stauffer1994introduction, Coniglio2016}.

%\sout{as the critical occupation probability $p_c$ is approached, suggesting an intimate connection with critical phenomena and, more in general, statistical physics.}

%\sout{ Several authors have proposed a description of  thermodynamic phase transition and nucleation based on the geometrical definition of clusters of particles. One of the first attempt was the theory of condensation  for a fluid of particles, initially proposed by Mayer  %\cite{mayer:74} and further developed by Frenkel \cite{frenkel:200, frenkel:538}. This  of the system as  $PV=\sum_s n_s k_BT$, where $n_s$ is the  number of clusters of $s$ particles.  Clusters, assumed to be compact region of interacting particles, are treated as non interacting objects. }

In order to recover the 
%\sout{critical} 
behavior of 
%\sout{particles,}
a fluid near the liquid-gas (LG) critical point,
 Fisher  listed \cite{fisher_droplets} 
 %\sout{modified the theory listing}
  the properties that the ``right" 
clusters ({\it droplet})
%\sout{droplets}
must have 
%\sout{in proximity of the liquid-gas (LG) critical point} 
in order to represent thermodynamically correlated regions
\cite{stauffer1994introduction, Sator20031}: i) an infinite droplets is formed only at the critical point; ii) the order parameter of the phase transition is related to the size of the infinite droplet; iii) the compressibility (or susceptibility for magnetic systems) of the order parameter is proportional to the mean cluster-size; iv) the 
%\sout{statistical} 
thermodynamic
correlation length $\xi$, quantifying the spatial extent of thermal fluctuations, is proportional to the average radius of the clusters, 
i.e. the connectivity length
$\xi_c$.
%\sout{namelly the connectivity length. This picture} 

%\sout{The properties listed by Fisher guarantee that the droplets recover the expected scaling relations at the critical point. Nevertheless, Fisher did not give an operative definition  of the droplet and a functional form for how $p$ changes with $T$ and $P$. For example, Sykes and Gaunt shown that the definition of clusters as regions of n.n. occupied sites in  a lattice-gas, or as n.n. parallel spins in an Ising model, does not share the properties of the  Fisher's droplet, because it  gives $\xi_c>\xi$. This result can be easily proved at infinite temperature where $\xi=0$ while $\xi_c\geq 1$ (the probability of finding n.n. spins in the same state is 0.5 at infinite temperature) %\cite{Sykes1976} .}

The ``right'' cluster definition for Ising-like systems was 
%\sout{found  in 1969 and 1972} 
proposed in the '70s
 by Kasteleyn and Fortuin (KF). They shown that it is possible to map a ferromagnetic Potts model onto a corresponding percolating model \cite{Kasteleyn_Fortuin, Fortuin1972536}. 
Later, Coniglio and Klein (CK) \cite{coniglio1979,coniglioJPhysA1980} introduced the ``random site-correlated bond'' percolation ({\it site-bond correlated} percolation, for short)  to prove
 
%\sout{Using a site-bond correlated description of the Ising model, based on the KF theory,  Coniglio and Klein   \cite{coniglio1979,coniglioJPhysA1980}   (CK) proved}  
the equivalence between the thermodynamic and  
 %\sout{geometrical} 
percolation
 critical behavior. 
%\sout{ They shown that the critical point coincides with  $p_c$, that $\xi$ and $\xi_c$ diverge with same critical exponent $\nu$ and that the susceptibility  of a Potts model  and the mean cluster-size  of the site-bond correlated percolation  diverge with the same critical exponent $\gamma$. }
The CK approach is 
 at the base of the ``Cluster Monte Carlo'' dynamics 
introduced by Swendsen and Wang (SW) \cite{SwendsenPRL1987}, representing
 %\sout{that represented} 
 a major improvement in our ability to generate equilibrium configurations near a critical point and at lower temperatures \cite{binder_book}.
 
\subsection{Site-bond correlated percolation for the FS model} 

Following the KF and CK approaches, we 
%\sout{superimpose a percolation description on the water model. In general, it is possible to} 
map the physical system onto a  
site-bond correlated percolation model where bonds are set between n.n. variables  $\sigma$ with a probability that depends on their state, their interaction, $P$ and $T$, defining clusters of thermodynamically correlated degrees of freedom. Because in the FS model the $\sigma$-variables can interact with two coupling constants---$J$, between n.n. molecules, and $J_\sigma$, within the same molecule---we define two probabilities functions---$p_J$ and $p_{J_\sigma}$, respectively---to set bonds between, and within, molecules.

%\sout{percolation system as long as it is possible to find a mathematical function, depending on the Hamiltonian, that allows us to build clusters of statistically correlated degrees of freedom. It means that two interacting bonding variables  $\sigma$ have a certain probability, depending on the specific interaction, to belong to the same cluster. In our model the HB interaction is described by two terms, with coupling constants $J$ and $J_\sigma$, so we anticipate finding two probabilities, namely $p_J$ and $p_{J_\sigma}$. }
In order to 
%\sout{calculate}
find the functional forms for
 such probabilities
in the condensed phase, we observe that  the $\sigma$-part of the partition function, Eq. (\ref{eq2_bis},)
 in the liquid phase, where is $n_i=1$ $\forall i$,
 %\sout{and the clusters have $\xi_c>1$,}
 reduces to
\begin{equation}\label{eq2}
 \begin{array}{llll}
Z_{\{\sigma\}} & =  & \sum_{\{\sigma\}} &\prod_{\langle i,j \rangle} \left[ 1 + \left( e^{(J_{\rm eff}/k_BT)}-1\right) \delta_{\sigma_{ij},\sigma_{ji}} \right] \\\\
& & & \times  \prod_{i=1}^N \prod_{(k,l)_i} \left[ 1 +\left(e^{(J_\sigma/k_BT)}-1\right)\delta_{\sigma_{ik},\sigma_{il}}\right].
 \end{array}
\end{equation}

We now introduce the cluster definition and show how the partition function can be
equally 
expressed as a sum over all possible clusters 
%\sout{of}
associated to
a given configuration $\sigma$ \cite{chin-kun1984}. 
Following CK,
we define a cluster $C$ as a region of 
%\sout{$\sigma$} 
variables connected through 
%\sout{artificial} 
{\it fictitious}
bonds. The bonds do not affect the interaction energy 
%\sout{or the particles distribution {\it fictitious bonds}); they just}
but are introduced to
 define the connectivity between  
 %\sout{bonding variables $\sigma$,  independently on the specific   bonding  state of connected variables} 
the  variables \cite{coniglio1982}. 
  
In our coarse-grain model the degrees of freedom that are relevant for the HB formation are the  variables $\sigma$. We therefore, perform the percolation mapping
 %\sout{In our coarse-grain description, they They  do not depend on the average distance $r$ between n.n. water molecules  (as long as we are in the liquid phase), so we carry out such mapping only}
on 
%\sout{the term} 
the part of the partition function that depends on the bonding variables $\sigma$,
$Z_{\{\sigma\}}$.

For sake of simplicity, let's assume that we have only $J_\sigma$ interactions. Then we will generalize to the case including also $J$ interactions. 
We put fictitious bonds on $J_\sigma$ interactions with probability 
\begin{equation}\label{p_s}
 p_{J_\sigma}\equiv 1-e^{- J_\sigma/k_BT }
\end{equation} 
between two $\sigma$-variables in the same state and define a cluster as maximal set of $\sigma$-variables
 connected by bonds. We call these bonds $b_{J_\sigma}$.
We will show now that the definition in Eq.(\ref{p_s}) allows us to map the thermodynamic model into the percolation model.

\begin{figure}
\centering
 \includegraphics[scale=0.7]{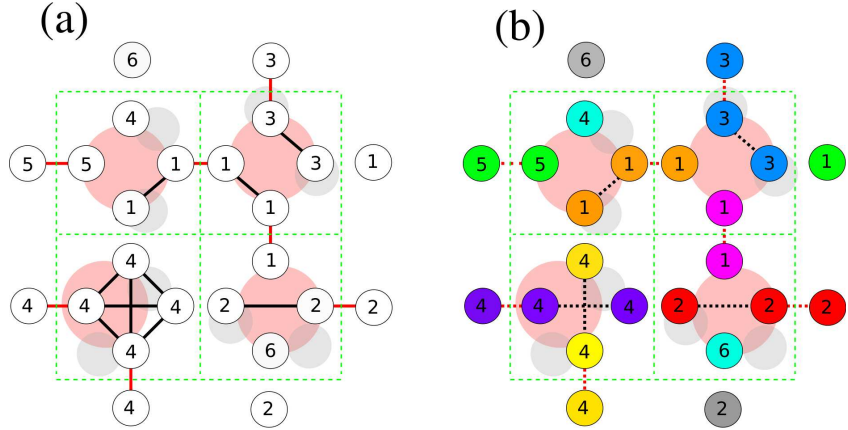}
 \caption{(a) Schematic drawing of the water model. The green lines identify the cell partition, containing a single water molecule. The molecule degrees of freedom are represented by four bonding variables $\sigma$, shown as white circles. Each variable $\sigma$ can assume up to $q=6$ different states (the numbers shown in each circle). The directional and cooperative interactions, with coupling constants $J$ and $J_\sigma$, between variables in the same state
are shown as red and black links, respectively, while interactions between variables in different states are not shown for sake of clarity.
(b) Schematic drawing of the percolation mapping. All the $\sigma$ belonging to the same cluster are depicted with the same color, and connected with ``fictitious" bonds. The fictitious bonds, shown with dotted lines, are set between two n.n. $\sigma$ in the same state with probability $p_J$ or $p_{J_\sigma}$ (given by Eq.s (\ref{p_j}) and (\ref{p_s}), respectively), according to the $J$ or $J_\sigma$ interaction. At finite $T$  
n.n. $\sigma$-variables in the same state have a finite probability to be disconnected and to belong to different clusters. In the example, n.n. $\sigma$-variables in the same state (e.g., 4 in the lower left) belongs to different clusters (colored in different colors), and the configuration has 
$|A|=19$, $|B|=5$, $|C|=5$, $|D|=14$, ${\cal N}=24$,
$|A_J|=5$, $|B_J|=0$, $|C_J|=7$, $|D_J|=5$, ${\cal N}_J=12$,
$N(C)=12$.
}
\label{scheme_mapping}
\end{figure}

First we observe that 
for a given configuration $\{\sigma\}$ 
we have several bond configurations that are ``compatible" with $\{\sigma\}$, i.e. such
that the fictitious bonds are set only between $\sigma$-variables in the same state. 
Let's call  $C$ one of these bond configurations,  
$|C|$  the number of bonds in the configuration $C$,
$|B|$  the number of ``missing" bonds,  i.e. the number of couples of $\sigma$-variables  in the same state without a bond, 
$|D|$  the number of couples of n.n. $\sigma$-variables in different states
(Fig. \ref{scheme_mapping}).

If we fix the configuration $\{\sigma\}$, then $|D|$  is fixed and
\begin{equation}
|C|+|B|={\cal N}-|D|
\label{1.34}
\end{equation}
where ${\cal N}$ is the total number of n.n. pairs of  $\sigma$-variables in the system.
Therefore,
for a given configuration $\{\sigma\}$ 
is 
\begin{equation}
\sum_Cp_{J_\sigma}^{|C|}(1-p_{J_\sigma})^{|B|}=1,
\label{1.35}
\end{equation}
because
\begin{equation}
\begin{array}{lcl}
\sum_C \!\!\!\!\!\!\!\!\!\! \raisebox{-9pt}{${\scriptstyle fixed \ |D|}$}~~
p_{J_\sigma}^{|C|}(1-p_{J_\sigma})^{|B|} & \equiv & \sum_{|C|=0}^{{\cal N}-|D|}
\left( \raisebox{8pt}{${\scriptstyle |C|+|B|}$}\!\!\!\!\!\!\!\!\!\!
\raisebox{-8pt}{${\scriptstyle |C|}$} ~~~\right)
p_{J_\sigma}^{|C|}(1-p_{J_\sigma})^{|B|}= \nonumber \\
 & = & [p_{J_\sigma}+(1-p_{J_\sigma})]^{{\cal N}-|D|}
\end{array}
\label{1.36}
\end{equation}
for the Newton binomial.
Therefore, we can write
\begin{equation}
\begin{array}{cl}
Z_{\{\sigma\}, J_\sigma} & \equiv \sum_{\{\sigma\}}
\prod_{i=1}^N \prod_{(k,l)_i} \left[ 1 +\left(e^{(J_\sigma/k_BT)}-1\right)\delta_{\sigma_{ik},\sigma_{il}}\right]
\sum_Cp_{J_\sigma}^{|C|}(1-p_{J_\sigma})^{|B|} \nonumber \\
 & =\sum_{\{\sigma\}}\sum_CW(\{\sigma\},C)
\label{1.38}
\end{array}
\end{equation}
where
\begin{equation}
W(\{\sigma\},C)=
\prod_{i=1}^N \prod_{(k,l)_i} 
\left[ 1 +\left(e^{(J_\sigma/k_BT)}-1\right)\delta_{\sigma_{ik},\sigma_{il}}\right]
p_{J_\sigma}^{|C|}(1-p_{J_\sigma})^{|B|}
\label{1.28}
\end{equation}
is the weight associated to the $\sigma$-variables configurations $\{\sigma\}$ together with the compatible bond configuration $C$.

If  we set
\begin{equation}
W(\{\sigma\},C)=0
\label{1.28bis}
\end{equation}
when $\{\sigma\}$ and $C$ are not compatible, 
then we can write
\begin{equation}
Z_{\{\sigma\}, J_\sigma} = \sum_C
\sum_{\{\sigma\}}W(\{\sigma\},C),
\label{1.38bis}
\end{equation}
where now the $\sum_C$ is done over all the bond configurations, independent of their compatibility with $\{\sigma\}$. 

For  n.n. $\sigma$-variables 
\begin{equation}
1 +\left(e^{(J_\sigma/k_BT)}-1\right)\delta_{\sigma_{ik},\sigma_{il}}
=\left\{ \begin{array}{ll}
1 & \mbox{if }  \sigma_{ik}=\sigma_{il} \nonumber \\
1- p_{J_\sigma} & \mbox{otherwise}   
\end{array}\right. 
\label{1.29}
\end{equation}
therefore,
\begin{equation}
W(\{\sigma\},C)= 
p_{J_\sigma}^{|C|}(1-p_{J_\sigma})^{|A|}
\delta_{\{\sigma\},C}
\label{1.28bis}
\end{equation}
where
$|A|\equiv |B|+|D|$ and 
\begin{equation}
\delta_{\{\sigma\},C}=\prod_{\langle k,l\rangle \in C}\delta_{\sigma_{ik},\sigma_{il}}
\label{1.31.b}
\end{equation}
takes into account the constraint on the compatibility of $\{\sigma\}$ and $C$.

If $N(C)$ is the number of clusters in the bond
configuration $C$, the number of configurations $\{\sigma\}$ compatible
with $C$ is $q^{N(C)}$, since every cluster can have only $q$
states with all the $\sigma$-variables equals.
Therefore, summing over $\{\sigma\}$ in Eq.(\ref{1.38bis})  and taking into account 
Eq.(\ref{1.28bis}) and Eq.(\ref{1.34}), we have
\begin{equation}
Z_{\{\sigma\}, J_\sigma} =\sum_Cp_{J_\sigma}^{|C|}(1-p_{J_\sigma})^{{\cal N}-|C|}q^{N(C)}.
\label{1.39}
\end{equation}

In the case in which we consider also $J$ interactions, we set fictitious bonds, called 
$b_J$, 
between n.n. $\sigma$-variables interacting with $J$ coupling  with probability 
\begin{equation}\label{p_j}
 p_{J}\equiv 1-e^{- J/k_BT }
\end{equation} 
and, generalizing the previous discussion and the notation in a straightforward way 
(Fig. \ref{scheme_mapping}), we can rewrite Eq.(\ref{eq2}) as
\begin{equation}
Z_{\{\sigma\}} =\sum_{C}
p_{J}^{|C_J|}(1-p_{J})^{{\cal N}_J-|C_J|} \times
p_{J_\sigma}^{|C|}(1-p_{J_\sigma})^{{\cal N}-|C|}q^{N(C)}
\label{1.39bis}
\end{equation}
where the cluster configuration $C$ includes bonds of both type $J$ and $J_\sigma$.
This expression of the partition function depends only on percolation quantities, completing the mapping between the thermodynamic system and the percolation problem.

Therefore, the mapping into the site-bond correlated percolation does not change the partition function of the model and its free energy. Furthermore, following FK/CK 
\cite{Kasteleyn_Fortuin, Fortuin1972536, coniglio1979,coniglioJPhysA1980},
it can be shown that this definition satisfies the Fisher's conditions for the ``right'' clusters, giving a geometric representation of  the thermodynamic droplets. 
As already mentioned, 
this observation allows us to adopt the well know SW Cluster Monte Carlo dynamics \cite{SwendsenPRL1987,PhysRevLett.62.361} 
that is able to generate equilibrium configurations in a very efficient way near phase transitions, where other Monte Carlo dynamics slow down and freeze \cite{binder_book}.

\section{Results}

\subsection{The thermodynamic behavior}

Based on the percolation description of the system it is possible to implement a cluster Monte Carlo algorithms \cite{PhysRevLett.62.361, binder_book} allowing faster equilibration in the deep supercooled region. In each Monte Carlo (MC) step we generate a cluster of fictitious bonds following the prescription of KF-CK-SW, with all the 
 variables
$\sigma_{ij}$ of the cluster in the same state by construction. 
Next, we
change at random the $\sigma$ state of the entire cluster, whose average 
linear
dimension $\xi_c$ is related to the correlation length $\xi$. Hence, approaching to a critical point, where $\xi$ becomes as large as the system size, the update involves a number of HBs of the order of the entire system, allowing a fast decorrelation of the new configurations and a more efficient MC sampling of independent configurations at equilibrium. This efficiency strongly contrasts with the {\it critical slowing down} that the local dynamics, such as Metropolis or Heat-Bath \cite{binder_book}, suffer near a critical point due to the increase of the correlation length $\xi$. 

 In the following  
we consider systems with increasing sizes, from $N=2.5\times 10^3$ up to $N=160\times 10^3$ water molecules, with periodic boundary conditions. We 
 perform annealing simulations at constant $P$ adopting
%\sout{simulate them along isobars}
the Cluster MC algorithm described above,
 following
%\sout{and in}
Ref.s  
\cite{Bianco2014, Mazza2009}.

In particular, our previous finite-size analysis of the model \cite{Bianco2014}  shown that the system is consistent with a thermodynamic LLCP for $P_Cv_0/(4\epsilon)\simeq 0.555\pm 0.002$ and $T_Ck_B/(4\epsilon)\simeq 0.0597\pm 0.0001$ with respect to a {\it mixed-field} order parameter $M=\rho^* + b'u^*$ \cite{PhysRevLett.68.193, PhysRevE.52.602, Wilding1996439}, given by a linear combination of  number density $\rho^*\equiv \rho v_0$ and energy density $u^*\equiv E/(4\epsilon N)$, 
where $b'$ is the mixing-coefficient defined in the Appendix, with $b'= 0.25 \pm 0.03$ for the explored range of system sizes\footnote{This 
definition of the order parameter, as shown in Ref.~\cite{PhysRevLett.68.193, PhysRevE.52.602, Wilding1996439}, allows to recover the 
 Isigin-like symmetry of the order parameter distribution, at the critical point, between the ordered and the disordered phases.}.
Such a critical point represents the ending point of a first order phase transition line with a negative slope in the $P$-$T$ plane, separating two metastable liquid phases, the HDL at higher $P$ and $T$, and the LDL at lower $P$ and $T$.
%\sout{Spanning} 
 Starting from the LLCP 
%\sout{toward lower $P$}
 we 
%\sout{have} 
found a line of maxima of $\xi$ 
 that extends toward negative pressures until reaching the stability limit of the stretched liquid.
%\sout{extending all the way down to the stability limit of stretched liquid at negative pressure, which} 
Following the definition given in the Appendix
we identified this line of $\xi$-maxima as the Widom line \cite{Bianco2014}. 
%In the following we present the results of the percolation analysis and compare them with the thermodynamic predictions. To ease the discussion of the results, we divide the section in subparagraphs for each specific topic discussed.

\begin{figure}
\leftskip 1 pt
{\bf (a) \hspace{5.4 cm} (b)}\\
 \includegraphics[scale=0.3]{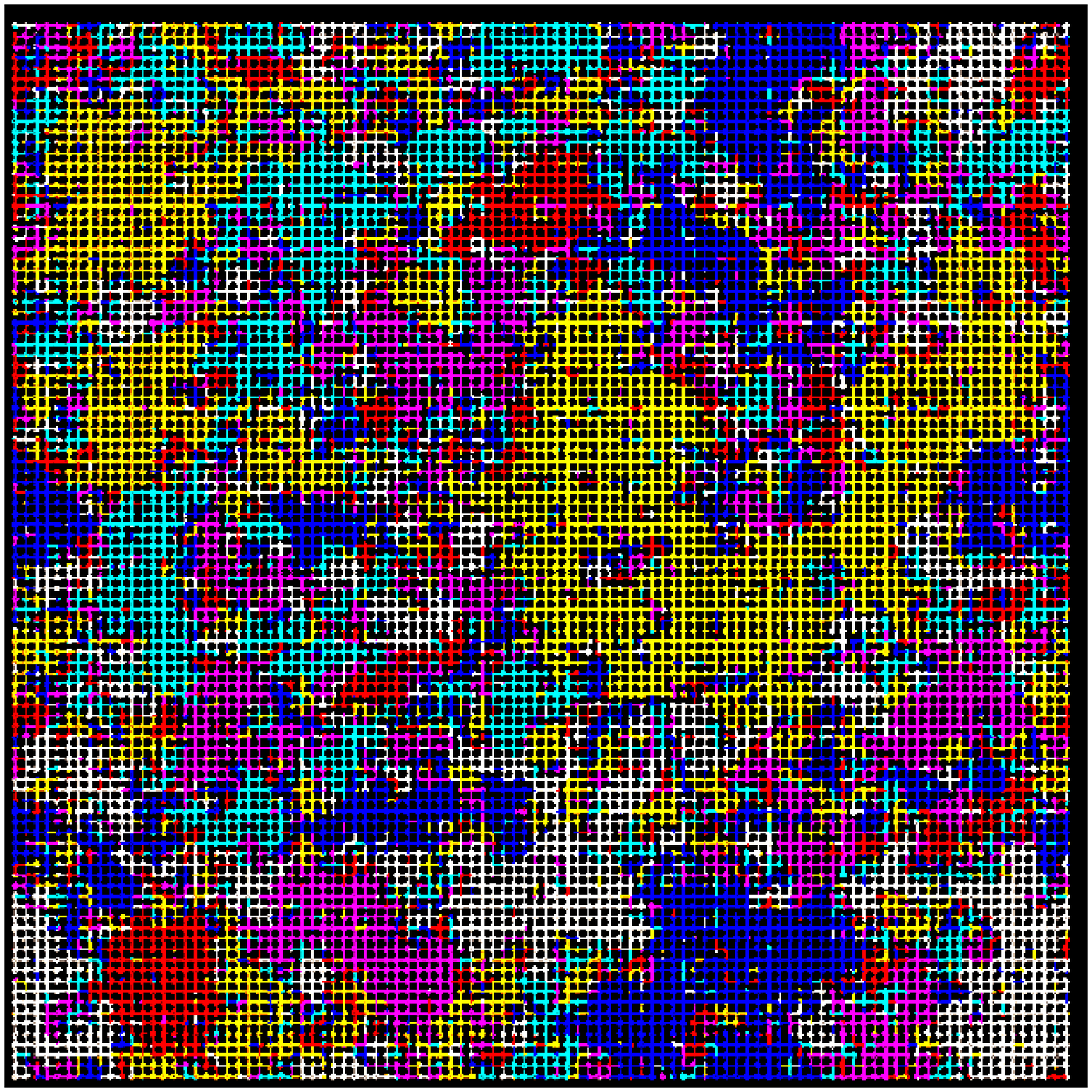}
\includegraphics[scale=0.3]{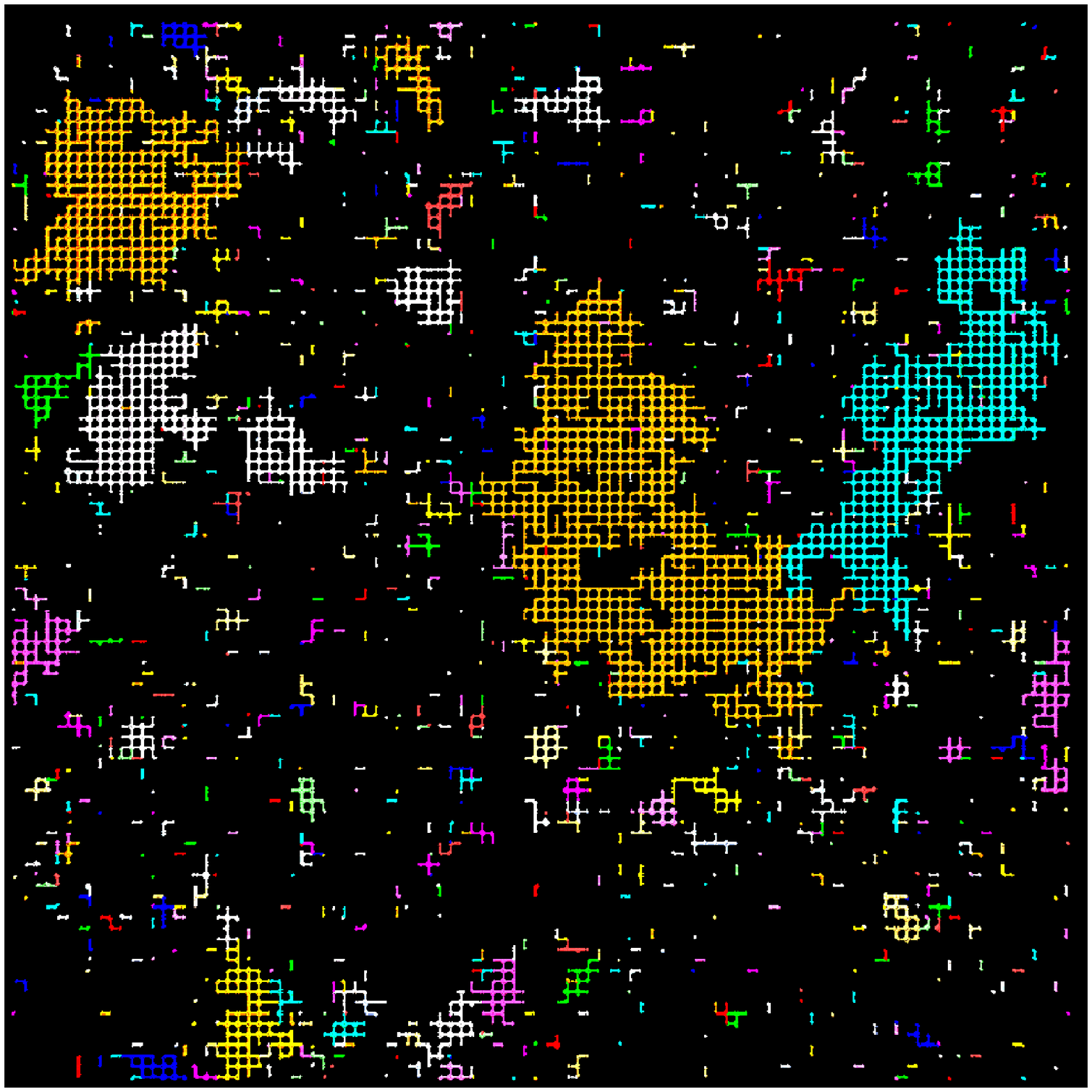}
\caption{{\bf (a)} A typical configuration of $\sigma$ variables for $N=10^4$ water molecules for $Pv_0/(4\epsilon)=0.6$ and $Tk_B/(4\epsilon)=0.058$. The state point is close to the LLCP at $P_Cv_0/(4\epsilon)\simeq 0.555\pm 0.002$ and $T_Ck_B/(4\epsilon)\simeq 0.0597\pm 0.0001$ \cite{Bianco2014}. In the high resolution image, $\sigma$ variables are represented by points and  HBs by lines.  There are six colors, one for each of the six possible $\sigma$ states.  
{\bf (b)} 
A possible cluster configuration
%\sout{Clusters}
 of correlated $\sigma$ variables. From the configuration in panel (a), we select only those variables in the yellow states (all the others are represented here in black) and among them, according to the probabilities in Eqs. (\ref{p_s}) and (\ref{p_j}), we build the clusters of correlated variables, representing
%\sout{them with the same}
each separate cluster with a different color. The resulting clusters have different sizes $s$, each representing a thermodynamically correlated region of H-bonded water molecules.
All the clusters in panel (b) are finite.}
\label{cl}
\end{figure}

\subsection{Percolation probability.} 

\begin{figure}
{\bf (a) \hspace{5.4 cm} (b)}\\
\includegraphics[scale=0.3]{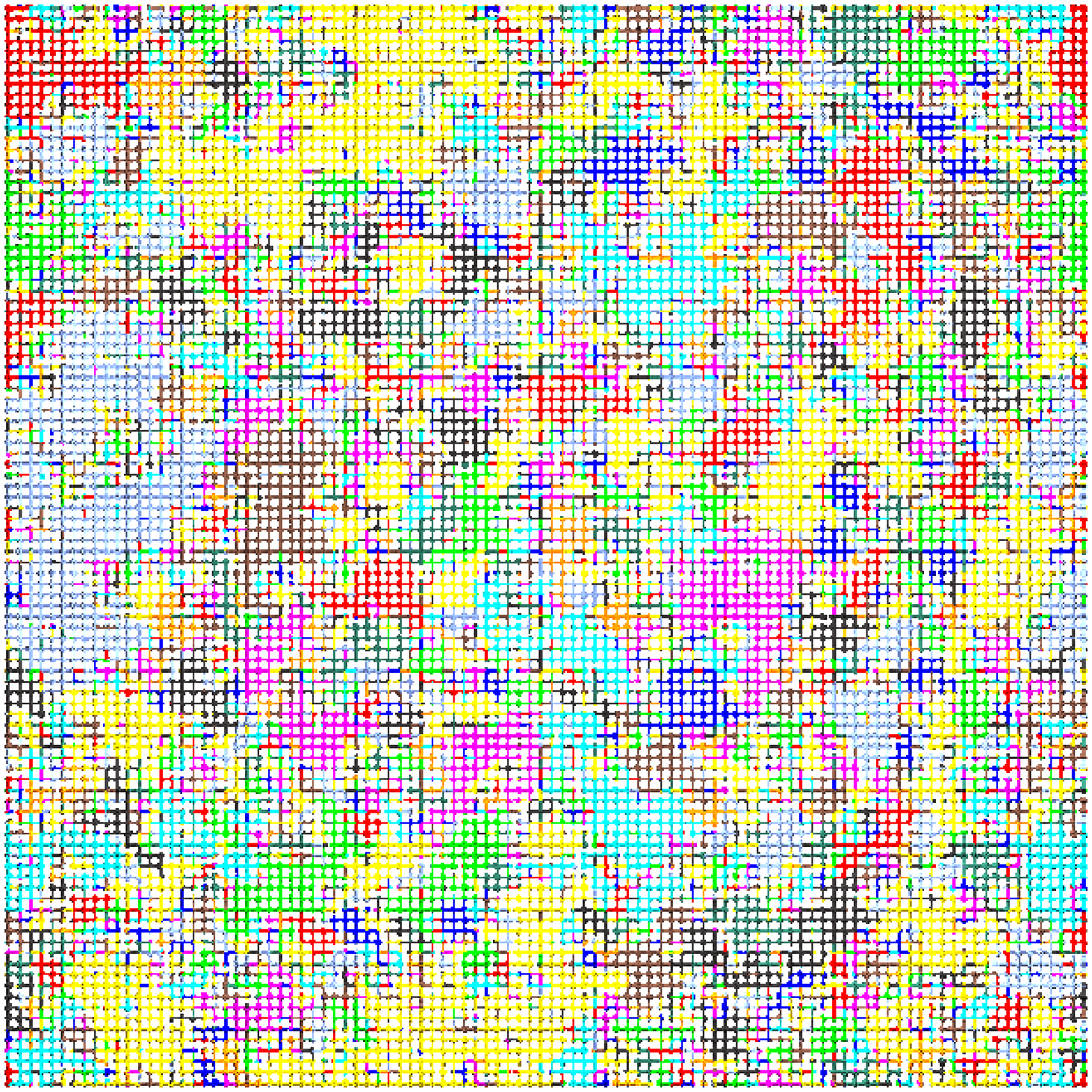} %\hspace{1 mm}
\includegraphics[scale=0.3]{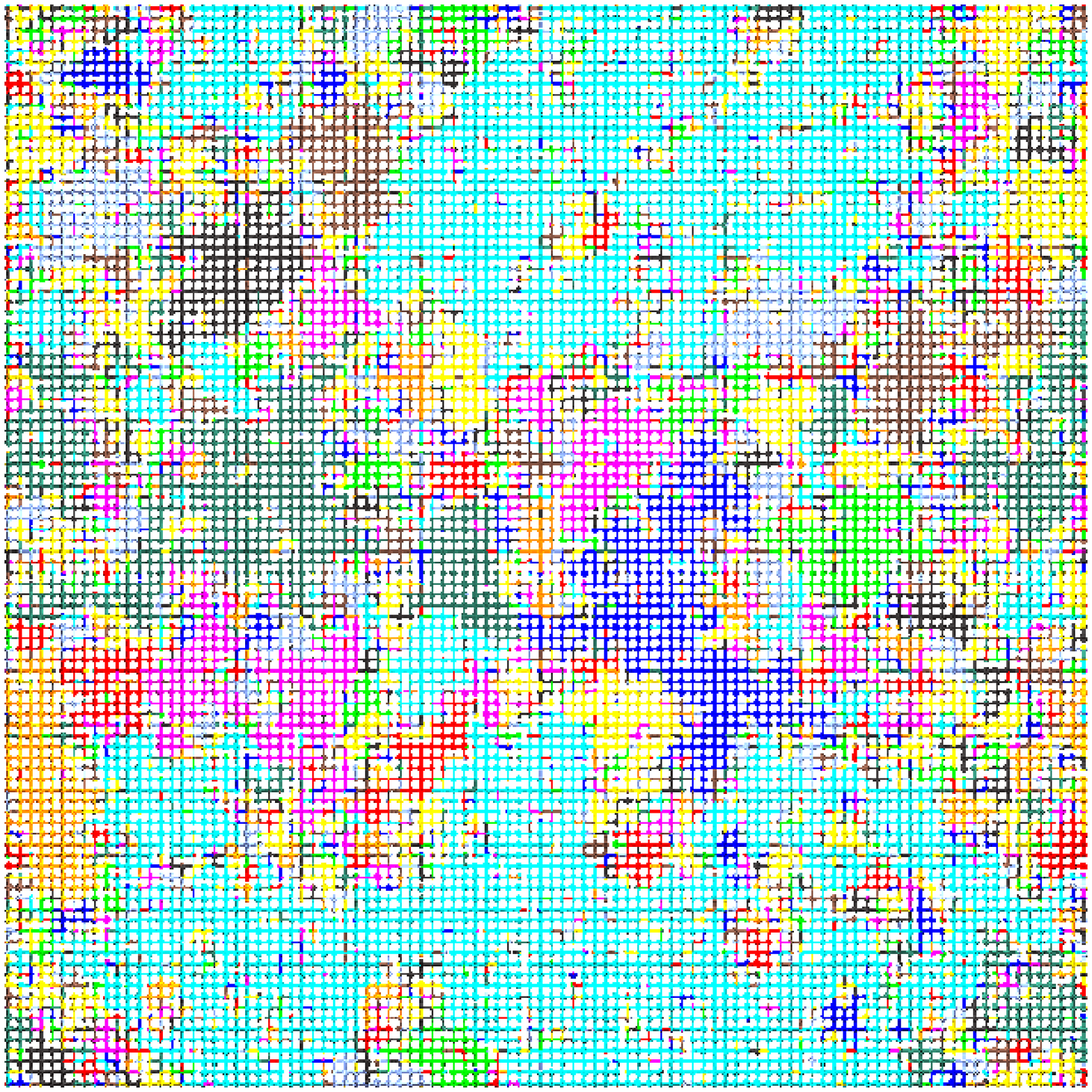} %\hspace{1 mm}
{\bf (c) \hspace{5.4 cm} (d)}\\
\includegraphics[scale=0.3]{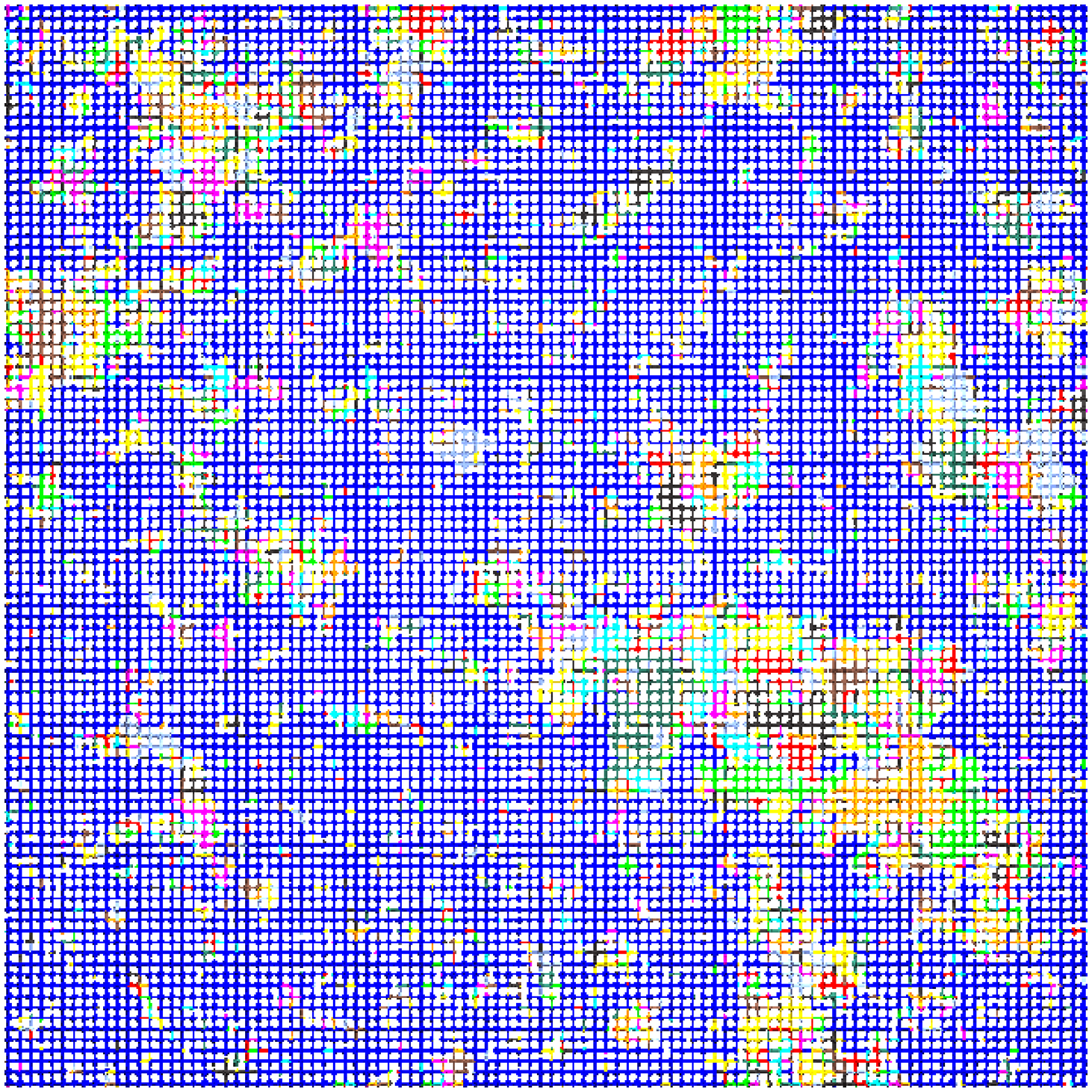} %\hspace{1 mm}
\includegraphics[scale=0.3]{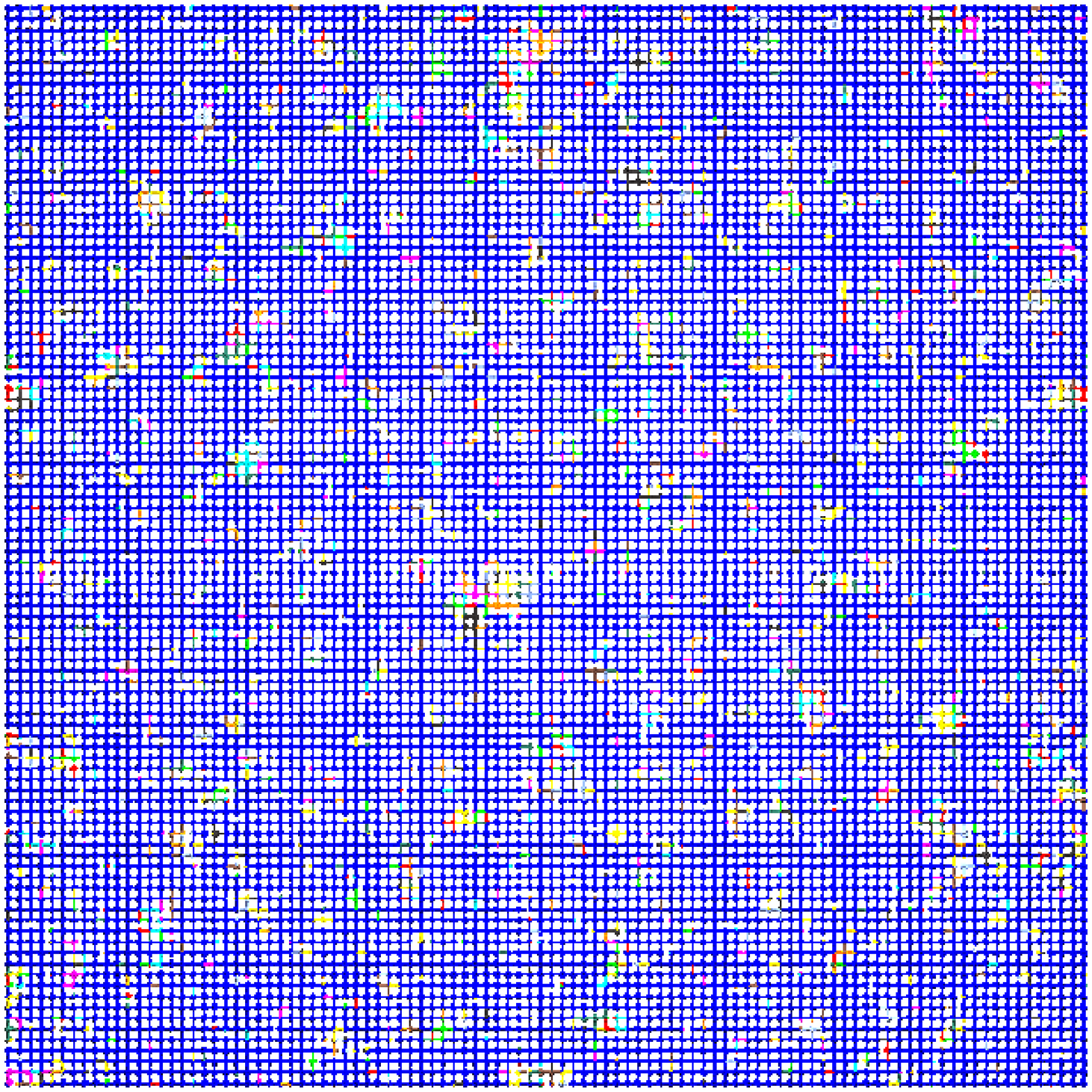}
\caption{Cluster configurations  of correlated $\sigma$-variables for $N = 10^4$ water molecules 
 at the critical isobar $P_Cv_0/(4\epsilon)\simeq 0.555$. As in  Fig. \ref{cl}, clusters can have six different states (colors). Data are 
%\sout{HBs are colored with six different colors corresponding to the six possible different states of the bonding variables $\sigma$. Furthermore, connected HBs with the same color belong to correlated water molecules. Data are taken along the critical isobar}
for:  {\bf (a)} $Tk_B/(4\epsilon)=0.058360$ (slightly above the percolation threshold); {\bf (b)} $0.058354$ (percolation threshold for the finite system, close to the LLCP temperature extrapolated for an infinite system $T_Ck_B/(4\epsilon)\simeq 0.0597$ \cite{Bianco2014}); {\bf (c)} $0.058340$ (slightly below the percolation threshold); {\bf (d)} $0.057000$. At the percolation threshold (b) we observe a cluster (turquoise) spanning the system from one side to the opposite, while above (a) and below the threshold (c, d) finite clusters are much smaller than the linear size of the system. Below the percolation threshold (c, d) there is always an infinite cluster (blue).}
\label{transitionB}
\end{figure}

The size $s$ of a cluster is defined as the number of $\sigma$ variables belonging to it.
% Each four $\sigma$ variable, on average, an entire water molecule belongs to a cluster. 
In Fig. \ref{cl} we show an example of clusters 
configuration
for a specific 
 $\sigma$-variables
configuration. The occurrence of a percolation transition is marked by the appearance of
 an {\it infinite} cluster spanning the entire system, i.e. a cluster with a linear size $\xi_c$ comparable or equal to the system's size (Fig.\ref{transitionB}).
%\sout{a percolating cluster, i.e. a clusters connecting two opposite sides of the system. }
The probability that an arbitrary $\sigma$-variable is part  of a 
%\sout{non-percolating}
 finite
 cluster of size $s$ is $\mathscr{P}_s\equiv n_ss$, where $4Nn_s$ is the average number of 
 %\sout{non-percolating} 
 finite
 clusters of size $s$ 
 %\sout{per site}
 , i.e. the fraction of 
 %\sout{the system that belongs}
 $\sigma$-variables that belong to clusters of 
 %\sout{the given}
 size $s$.
 Therefore, the probability that an arbitrary $\sigma$-variable belongs to {\it any} 
 %\sout{non-percolating} 
 finite
 cluster is $\mathscr{P}_{<\infty}\equiv \sum_s \mathscr{P}_s$, and 
its probability to belong to the 
%\sout{percolating} 
 infinite cluster, i.e. the {\it percolation probability} for 
 %\sout{a generic}
 any
   $\sigma$-variable, is \cite{stauffer1994introduction} 
\begin{equation}\label{perc} 
 \mathscr{P}_\infty \equiv 1-\mathscr{P}_{<\infty} .
\end{equation}

\begin{figure}%[h!]
\leftskip 10 pt 
{\bf (a)}\\
\centering
\includegraphics[scale=1.3]{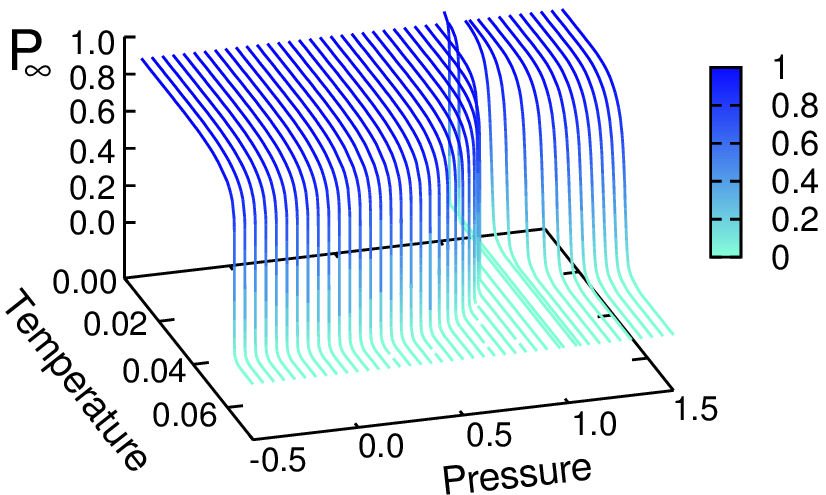}\\
\vspace{0.2 cm}
\leftskip 26 pt 
{\bf \hspace{1 cm}(b)\hspace{2.6 cm} (c) \hspace{2.5 cm} (d)}\\
\vspace{0.1 cm}
\centering
\includegraphics[scale=0.4]{percolation_sizes.eps}
\caption{{\bf (a)} Percolating probability $\mathscr{P}_\infty$,  for $N=10^4$ water molecules, along  isobars for $-0.5\leq Pv_0/(4\epsilon) \leq 1.5$ and $0\leq Tk_B/(4\epsilon)\leq 0.08$, corresponding to the supercooled region of liquid water,  from stretched water (negative pressure) to very high pressures ($Pv_0/(4\epsilon)>1$). The color code along lines represents the value of $\mathscr{P}_\infty$. Each line joins $\sim 150$ simulated state points  for each value of $P$, with a statistic of $\sim 10^5$ independent configurations. 
Finite size effect for $\mathscr{P}_\infty$ for $N$ between $2.5\times 10^3$ (black circles) and $108.9\times 10^3$  (orange left triangles) as function of $T$ for 
{\bf (b)} $Pv_0/(4\epsilon)=0.3$, 
{\bf (c)} $Pv_0/(4\epsilon)=0.9$,
{\bf (d)} $Pv_0/(4\epsilon)=1.5$.
In each panel, $P$ and $T$ are expressed in units of $v_0/(4\epsilon)$ and $k_B/(4\epsilon)$, respectively.}
\label{p}
\end{figure}

The quantity  $\mathscr{P}_\infty$ represents the order parameter for the percolation transition, 
 with
%\sout{We have}
$\mathscr{P}_\infty = 0$ when there is no 
%\sout{percolating cluster} 
 percolation  and $0< \mathscr{P}_\infty \leq 1$ otherwise
%\sout{$\mathscr{P}_\infty = 1$ when there is  probability 1 of finding a percolating cluster } 
(Fig. \ref{p}.a). 
For finite systems $\mathscr{P}_\infty$ increases rapidly near the percolation transition, with a larger slope for larger systems (finite size effect, Fig. \ref{p}.b,c,d.).
By convention, for any finite size the percolation threshold is at the thermodynamic point where $\mathscr{P}_\infty = 0.5$.
 
Our numerical findings reveal the occurrence of a percolation transition in all the range of simulated pressures. 
According to $\mathscr{P}_\infty$ we can distinguish at least three regions \footnote{This separation in three regions is reinforced by the analysis of the cluster-size distribution presented  elsewhere.
%\sout{in the following.}
}. 
The first region corresponds to $Pv_0/(4\epsilon)\lesssim 0.5$, where the percolation transition is sharp, becoming sharper for larger  $N$, with an evident finite-size effect for $N<25.6\times 10^3$  (Fig. \ref{p}.b).  
The second region is for $0.6<Pv_0/(4\epsilon)<1$, where  
$\mathscr{P}_\infty$ increases sharply at the percolation threshold and the finite-size effect is weaker than at lower $P$, being more evident only for $N< 10\times 10^3$  
%goes from 0.1 to 0.7 in a $Tk_B/(4\epsilon)$-range of $\sim 2\times 10^{-4}$ for $N=2500$, while the range is $<10^{-5}$ for $N=108900$ 
(Fig.s \ref{p}.c). 
The third region corresponds to $Pv_0/(4\epsilon)>1$, where the percolation transition is smoother for any system size $N$ (Fig. \ref{p}.d). 

\begin{figure}%[h!]
\centering
\includegraphics[width=1\textwidth]{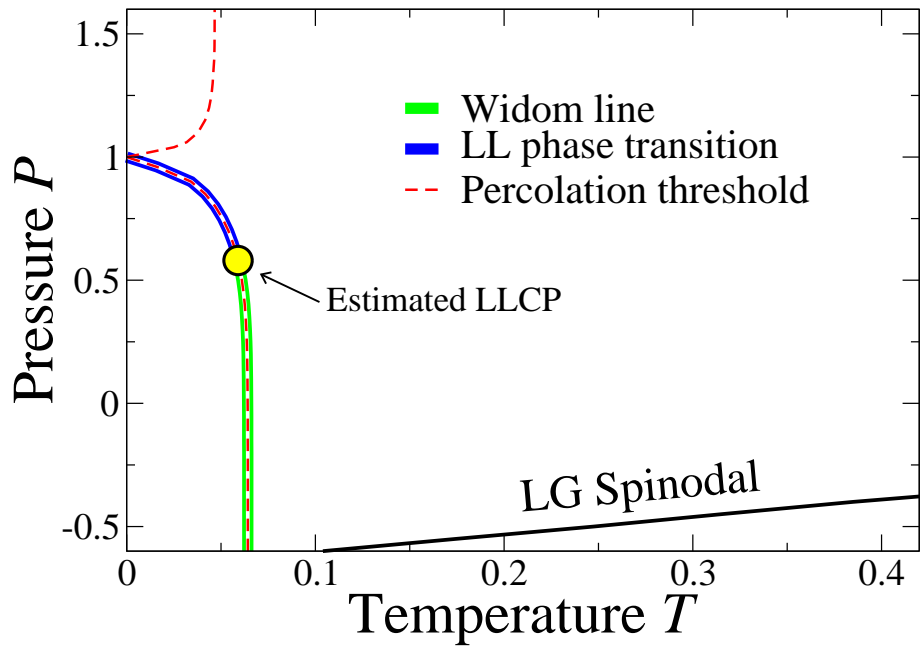}
\caption{Percolation lines in the $T-P$ plane, for $N=10^4$, compared with the LLPT
and the Widom line calculated with thermodynamic analysis in  Ref. \cite{Bianco2014}. The yellow circle identifies the thermodynamic LLCP and the continuous black line at negative $P$ is the liquid-gas spinodal. $T$ and $P$ are in units of $4\epsilon/k_B$ and $4\epsilon/v_0$, respectively.
}
\label{clust_distA}
\end{figure}

%\sout{By defining the percolation threshold where $\mathscr{P}_\infty=0.5$, we locate the locus of percolation line in the $P$--$T$ plane (Fig. \ref{clust_distA}). We find}
 Our analysis reveals
two percolation lines in the thermodynamic plane ($T$,$P$) emanating from the state point (0,1), expressed 
 in units 
$4\epsilon/k_B$ and $4\epsilon/v_0$, respectively (Fig. \ref{clust_distA}). 
The first line, for $Pv_0/(4\epsilon) < 1$, 
has a negative slope that decreases for decreasing $P$ and increasing $T$, becoming almost $T$-independent for sufficiently low $P$. This percolation line coincides, according to the previous analysis \cite{Bianco2014}, with the 
 first-order
LLPT for $0.55\lesssim Pv_0/(4\epsilon)<1$, while for 
%\sout{lower} 
 $Pv_0/(4\epsilon)\lesssim 0.55$
it coincides with the locus of strong maxima of $C_P$, $K_T$ and thermal expansivity $\alpha_P$. Furthermore, in Ref. \cite{Bianco2014} we shown that the locus of strong maxima of the response functions is also where the thermodynamic correlation length $\xi$ of the HBs has a maximum. Hence, 
%\sout{it}
the percolation line at low $P$
 corresponds to the  Widom line.

The second percolation line is observed for $Pv_0/(4\epsilon)>1$ 
and has a positive slope that increases for increasing $P$ and $T$
 (Fig. \ref{clust_distA}). It  coincides approximately with the thermodynamic 
 %\sout{loci}
locus 
of weak maxima of $C_P$ along isotherms
%\sout{weak maxima of $K_T$ along isobars and weak minima of $\alpha_P$ along isotherms}
that  we found at high $P$ \cite{Bianco2014}.
 As we have shown \cite{Bianco2014}, this locus coincides also with that of weak maxima of $K_T$ along isobars and with that of weak minima of $\alpha_P$ along isotherms, consistent with similar lines found in other  models with the LLCP \cite{Luo2014, Abraham2011, Buldyrev:2015uq, Luo2015}.

\begin{figure}%[h!]
\includegraphics[scale=0.49]{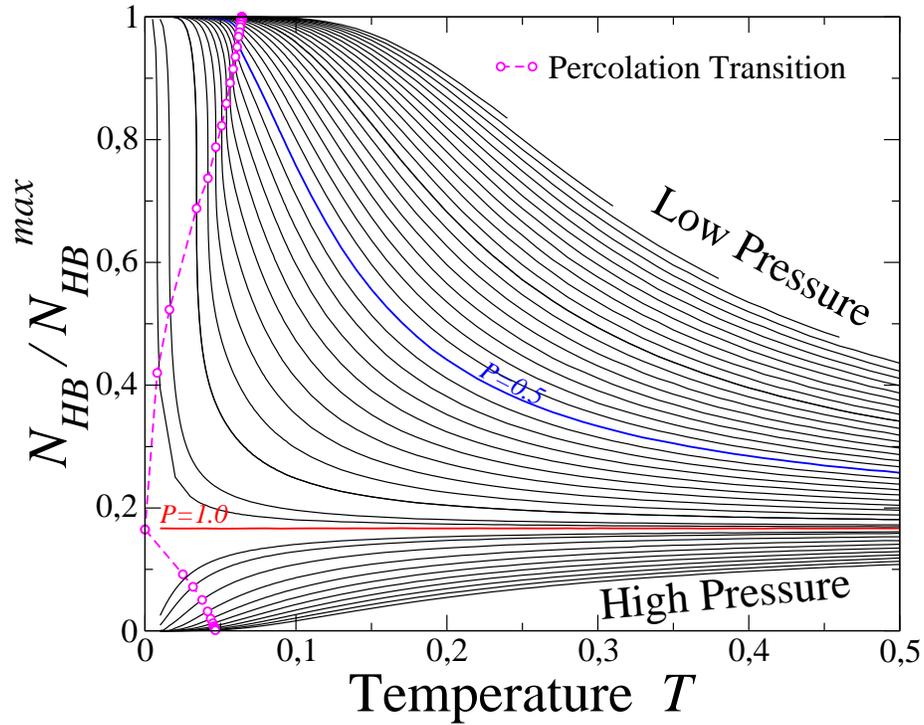}
\caption{Number of HB $N_{HB}$, normalized to its  maximum 
 per molecule
$N_{HB}^{\rm max}\equiv2N$, as function of $T$ along isobars for $N=10^4$ water molecules. Black lines join simulated state points ($\simeq 150$ for each isobar).  The magenta circles mark the percolation temperature along the isobars. The blue and red lines emphasize the $N_{HB}$ along the isobars $Pv_0/(4\epsilon)=0.5$ and $1.0$, respectively.
Pressures are, from top to bottom, from $Pv_0/(4\epsilon)=-0.60$ to 1.55 in steps of 0.05, apart from one extra isobar at $Pv_0/(4\epsilon)=0.975$.}
\label{hb}
\end{figure}

\section{Discussion} 

\subsection{Hydrogen bond network and the percolation line.}

%Next we discuss the relation between the percolation transition and the building up of the HB network. 
For our choice of parameters and $Pv_0/(4\epsilon)\lesssim 1$, i.e. for $P$ such that $J_{\sigma}\lesssim J_{\rm eff}$,  the percolation line marks the threshold for the formation of a network of correlated HBs. In this range of pressures, 
the number of HBs, $N_{\rm HB}$, increases monotonically  and saturates to its maximum value upon cooling \cite{Mazza2012}. The  increase is smooth at low $P$ and becomes sharper approaching $Pv_0/(4\epsilon)=1$  (Fig. \ref{hb}).

Mazza et al. shown that in this range of $P$ the number of cooperative HBs, i.e.  those HBs that become correlated, sharply increases at $Tk_B/(4\epsilon)\simeq 0.05$, with a locus of maximum-derivative that coincides with (i) the LLPT at  $P>P_C$, (ii) the LLCP 
 at the critical pressure, and (iii) the  locus of strong maxima of $K_T$ (indicated as $K_T^{\rm coop}$ in Ref. \cite{Mazza2012}) at $P<P_C$. Here,  these loci coincide with the percolation threshold for $Pv_0/(4\epsilon)\leq 1$ (Fig.\ref{clust_distA}), as we discuss next.

%In particular for $1< P \lesssim 0.5$ the clusters percolate when $N_{\rm HB}$ increase sharply. 
%The jump of $N_{\rm HB}$ points out an abrupt decrease of density, due to the thermodynamic phase transition between HDL and LDL.
%By comparing the locus of the percolation transition and the loci of the thermodynamic response functions reported in \cite{Bianco2014}, we conclude that, for the considered range of pressures, the percolation transition overlaps with the line of strong extrema of $C_P$, $K_T$ and $\alpha_P$. (Fig. \ref{hb}). 

For $P<P_C$, the HB network is gradually formed and the percolation threshold occurs when the system is highly H-bonded, i.e. $N_{\rm HB}\simeq 95\% N_{\rm HB}^{\rm max}$ (Fig. \ref{hb}). 
 
Nevertheless, at $T$ above the percolation line the clusters of H-bonded molecules are made of isolated HBs, i.e. the clusters are made, mainly, of $b_J$ bonds on $J$ interactions. On the other hand, 
at $T$ below the percolation line these clusters coalesce into clusters that span the entire system as a consequence of the setting of a macroscopic number of $b_{J_\sigma}$ bonds on $J_\sigma$ interactions.
This setting implies local rearrangements of the $\sigma$-variables toward more ordered configurations under the action of the $J_\sigma$ interactions, i.e. a local reordering of the HB network due to  the many-body interaction among the water molecules. 

%\sout{At this thermodynamic conditions,  we can associate the spanning clusters to a local reordering of the HB network }
The local reordering of the molecules does not affect the density of the system, but involves large fluctuations in energy and entropy. As a consequence, 
the percolation line coincides with the loci of strong maxima in isobaric specific heat, isothermal compressibility and isobaric heat expansion coefficient previously reported in Refs. \cite{Mazza2012, Bianco2014}.

The comparison of the percolation line for $Pv_0/(4\epsilon)<1$ (Fig. \ref{clust_distA}) with the locus of maxima of $\xi$, reported in Fig.6.a,b of Ref. \cite{Bianco2014}, reveals that both loci coincides within the numerical error. Hence, 
%on the one hand, 
this confirms that our percolation mapping is capable to capture %\sout{, through the proper definition of clusters,}
 the statistical fluctuations of the water molecules 
%\sout{(further data fostering the thermodynamic-percolation mapping will be discussed in the next subparagraph).} 
%On the other hand, our new findings could suggest that the percolation line at low $P$, overlapping with the locus previously identified as the Widom line \cite{Bianco2014}, reveals instead the presence of a critical thermodynamic transition induced by the confinement. 

%\sout{It is worth observing here that}
The mapping between the percolation problem and the thermodynamic system strictly holds only at a critical transition, because the cluster definition assumes a diverging correlation length at the percolation threshold. Therefore, from the percolation mapping we can draw useful conclusions for the thermodynamic system only along the critical isobar.
Nevertheless, the coincidence of the percolation line with the LLPT, above $P_C$, and the loci of strong fluctuations, below $P_C$, shows that the clusters of correlated HBs play a role in a large range of pressures, even if their spanning is not associated to critical fluctuations of thermodynamic quantities. 

This observation reconciles two classes of thermodynamic scenarios that have been debated as possible explanations  of the anomalous water behavior.
%\sout{, those assuming}
The first class assumes
the occurrence of a LLPT 
and includes two scenarios: the {\it LLCP-scenario} \cite{llcp} and the {\it LLCP-free scenario} \cite{Angell2008}.
 The second class includes the {\it Singularity-free scenario} \cite{Sastry:1996aa} that can be interpreted
%\sout{and that}
assuming a percolation behavior  \cite{Stanley:1979aa, Stanley:1980aa}. 
Here, as in Refs. \cite{Stanley:1979aa, Stanley:1980aa}, the correlation length $\xi$ increases as $T$ decreases as a consequence of the increasing size of clusters made of correlated water molecules. However,
the correlated percolation approach presented here is different from the one proposed in the ``polychromatic correlated-site percolation" \cite{Stanley:1979aa, Stanley:1980aa}, in which tiny ``patches", made of water molecules with all HBs formed, are of lower density than the rest of the HB network and 
%\sout{hence} 
give rise to anomalous density fluctuations. Furthermore, here we give the bond probability as an explicit function of $T$ and $P$, while in Refs. \cite{Stanley:1979aa, Stanley:1980aa} the authors give only qualitative relation between the percolation quantities and the thermodynamic observables, leaving the determination of the functional form to the experiments. 
%\sout{However, both approaches lead to an increase of $\xi$ as $T$ decreases as a consequence of the increasing size of clusters made of correlated water molecules.}

\subsection{The percolation at extremely high $P$} 

We find a percolation line, for our choice of parameters, also for $Pv_0/(4\epsilon) > 1$ (Fig. \ref{clust_distA}). At these pressures the percolation line is characterized by a smoother change in the order parameter $\mathscr{P}_\infty$ (Fig. \ref{p}d). 
The HB formation is enthalpically unfavoured, because $Pv_0>J$ and $J_{\rm eff}<0$. Hence, a decrease of $T$ induces HBs
 breaking with a decrease of $N_{\rm HB}$ (Fig. \ref{hb}).

 %\sout{At the same time} 
 Nevertheless, $J_\sigma>0$, hence
 the four $\sigma$-variable of the same water molecule assume the same state
 at low $T$, 
maximizing the probability to set bonds $b_{J_\sigma}$. At the same time, $J_{\rm eff}<0$ induces different $\sigma$-states on  n.n. water molecules. Under these conditions the  bonds $b_J$ have the maximum probability when the $\sigma$-variables have different states. 
 Hence, at $Pv_0/(4\epsilon) > 1$
 %\sout{while those of a n.n. water molecule tend to set in different $\sigma$-state to form a CK fictitius-bond. Hence,} 
 the clusters are made of anticorrelated water molecules. 
   It can be shown\footnote{Evidences will be provided elsewhere.} that under these conditions the percolation belongs to the class of the the {\it random-bond percolation} \cite{stauffer1994introduction} and that its percolation temperature
 %\sout{and the temperature of the random-bond percolation}
  increases for $Pv_0/(4\epsilon)>1$ as long as $|J_{\rm eff}|<J_\sigma$, i.e. up to $Pv_0/(4\epsilon)\simeq 1.1$, consistent with our finding (Fig.\ref{clust_distA}).
 
Interestingly, this random-bond percolation line could be identified with the ``Kert\'esz line'' or ``Coniglio-Klein line'' \cite{Kertesz198958, Adler1991222}
%, due to the KF percolation approach, 
extending from a percolation critical point at $T = 0$ and $Pv_0/(4\epsilon)=1$ up to $P = \infty$. 
%For example, at $T=0$ and $Pv_0/(4\epsilon)=1$ the system is frozen with all $\sigma$-variables in the same bonding state, as a consequence of the action of the 
% and the CK approach falls into a bond-percolation problem, undergoing a critical percolation transition \cite{stauffer1994introduction}. 
It is still under debate if the Kert\'esz line corresponds to a specific thermodynamic locus or not. For example,  it has been proposed that this line exists in the supercritical phase of a simple fluid \cite{Campi199746,Campi200124} and is related to the locus where vanishes the surface tension of droplets made of the denser fluid.

Finally, it is worth noting that the possibility of the presence of a second critical point in liquid supercooled water, at higher $P$ and lower $T$ with respect the  LLCP, has been predicted by Strekalova et al.  \cite{strekalovaPRL2011} in the case of water confined in a disordered matrix of hydrophobic nano-particles. Such a second CP could coincide with the critical percolation point at 
$T = 0$ and $Pv_0/(4\epsilon)=1$.\\

\section{Conclusions}  

In this work we  present a percolation study for a supercooled water monolayer adopting a many-body water model that has the LLCP \cite{Franzese:2002aa, FS-PhysA2002, Bianco2014}. The geometrical description of the system is based on the KF-CK-SW mathematical mapping \cite{Kasteleyn_Fortuin, Fortuin1972536, coniglioJPhysA1980, PhysRevLett.62.3054, SwendsenPRL1987} of the correlated site-bond percolation
that allows us to compute clusters of correlated degrees of freedom along the critical isobar of the LLCP. In such a way, at $P_C$ (i) the average size of a cluster coincides with the thermodynamically correlated regions, (ii) the cluster connectivity length $\xi_C$ coincides with the thermodynamical correlation length $\xi$, and (iii) the percolation set of critical exponents belongs to the same universality class of the thermodynamic critical exponents. 

The percolation description can be extended to any $P$, although the mapping  with the thermodynamically correlated regions at $P\neq P_C$ is not exact. The discrepancy comes from the fact that the percolation phenomena is critical, by construction, at any $P$, while the thermodynamic system not necessarily is. Hence at any $P$ we always reach a percolation threshold with a diverging $\xi_C$, even where $\xi$ is not critical. Nevertheless, the percolation approach reveals interesting results not only at $P_C$.

In particular, 
we find  a line of percolation transitions, with negative slope in the $P$--$T$ plane, 
that starts at $Pv_0/(4\epsilon)=1$ and $T = 0$ 
and extends up to the liquid-gas spinodal at negative pressures and $T>0$.
This line coincides with the loci of strong extrema of the specific heat $C_P$, compressibility $K_T$ and thermal expansivity $\alpha_P$, as resulting from previous thermodynamic analysis \cite{Mazza2012, Bianco2014}.
For pressures above $P_Cv_0/(4\epsilon)\simeq 0.6$ and below $Pv_0/(4\epsilon)=1$, the percolation transition coincides with the thermodynamic LLPT  separating two liquids with different energy and density \cite{Stokely2010, Mazza2012, Bianco2014}.

%\sout{We calculate the  cluster-size distributions along the percolation line and find that it decays with power laws,  whose characteristic exponent $\tau$  increases as $P$ decreases. As a consequence, the cluster fractal dimension is $D\simeq 2$ at $Pv_0/(4\epsilon)=1$ and is $D\simeq 1.63$ for $Pv_0/(4\epsilon)\lesssim 0.5$, becoming  less compact at lower pressures.  }

Near $P_C$ the critical behavior of the site-bond correlated clusters
%\sout{belongs to the 2d Ising universality class,} 
 is
consistent with our thermodynamic analysis \cite{Bianco2014}
%\sout{This result confirms the correctness of the percolation mapping at $P_C$, where}
and
the appearance of the spanning cluster corresponds to a cooperative reorganization  of the network of correlated HBs.
At lower pressures, $Pv_0/(4\epsilon)\lesssim0.5$, 
%\sout{at $T$ below the percolation transition the occurrence of a significant number of small clusters possibly reveals the heterogeneity of the system. At these low pressures we cannot claim that the percolation mapping is correct, because no critical behavior has been revealed by the thermodynamic analysis so far. Instead} 
we observe the coincidence of the percolation line with the Widom line  estimated from the direct calculation of isobaric maxima of $\xi$ \cite{Bianco2014}. Hence, at low pressures the cluster size 
%\sout{seems to increase} 
 increases
more than the correlated regions, because $\xi_C$ diverges while $\xi$ 
%\sout{does not. However, $\xi$}
only 
has a maximum 
 %\sout{where $\xi_C$ diverges, hence  the percolation threshold marks}
 marking
 the Widom line. 

Furthermore, a modified version of the model for the water monolayer, developed by Vilanova and Franzese and including crystal phases and water polymorfism \cite{Vilanova2011}, reveals the occurrence of a hexatic phase separated from the liquid by a critical line in the deep supercooled region. This finding has been confirmed by molecular dynamic simulations of  an atomistic model for a bilayer of water \cite{Zubeltzu:2016aa}. It is, therefore, interesting to analyze if the percolation transition could be eventually related to the hexatic critical line at low $P$ in the extended model. We are currently investigating this possibility. 

Our percolation approach leads, instead, to a different description at pressures above the LLPT. 
Emanating from the state point at $T = 0$ and $Pv_0/(4\epsilon)=1$, we find a percolation line with positive slope, extending up to infinite pressure,  in the $P$--$T$ plane. This line is  characterized by a smooth percolation transition 
%\sout{and our analysis shows that it corresponds to}
as in
 the random-bond percolation 
 %\sout{, as expected.}
and
 %\sout{The percolation line at $Pv_0/(4\epsilon)>1$}
approximately corresponds to the loci of weak extrema found for $K_T$, $C_P$ and $\alpha_P$ at these pressures in this model \cite{Bianco2014}, 
 %\sout{and} 
 consistent with similar lines found in other models with the LLCP \cite{Luo2014, Abraham2011, Buldyrev:2015uq, Luo2015}.

An intriguing hypothesis is that this percolation line at $Pv_0/(4\epsilon)>1$ 
could be identified with the Kert\'esz line \cite{Kertesz198958}, emanating from the percolation point at $T = 0$ and $Pv_0/(4\epsilon)=1$. Similar lines have been 
  observed in the Ising model around the Curie point and in Lennard Jones systems around the LG critical point \cite{Sator20031}. Such line, that in our model marks the decrease of the 
number of HBs,
  $N_{\rm HB}$, at very high pressure,
could be related to the locus where vanishes the surface tension of droplets made of the denser fluid  \cite{Campi199746,Campi200124}. Further investigation, beyond the scope of this work,  is necessary to understand this possibility.

\section*{Acknowledgments}

V. B. acknowledges the support from the Austrian Science
Fund (FWF) grant No. M 2150-N36, and from the the European Commission through the Marie Sk\l{}odowska-Curie Fellowship No. 748170 ProFrost.
G. F.  acknowledges support from the ICREA Academia prize, funded by the Generalitat de Catalunya and from Spanish MINECO grant
%FIS2012-31025 and 
FIS2015-66879-C2-2-P.

\section*{Appendix}
According to the scaling theory, and following  Refs.~\cite{Holten2012a}, \cite{Luo2014},
%{\it V. Holten et al., J. Chem. Phys. 136, 94507 (2012)} and {\it J. Luo, et al., Phys. Rev. Lett. 112, 135701 (2014)}, 
the critical behavior of a fluid can be described in terms of two independent scaling fields: the ordering (strong) field $h_1$ and the thermal (weak) field $h_2$
and the Widom line is identified as the locus in the $T$-$P$ plane where $\xi$ has a maximum, calculated/measured along the path $h_2=$constant.

This can be seen by considering that a generalized expression for the free energy is given by the scaling field $h_3$, depending on $h_1$ and $h_2$, that close to the critical point is
\begin{equation}
 h_3(h_1,h_2) \approx |h_2|^{2-\alpha} f^{\pm} \left ( \dfrac{h_1}{|h_2|^{\gamma + \beta}} \right)
\end{equation}
where $\alpha$, $\beta$ and $\gamma$, related to each other by the scaling relation $\alpha + 2\beta + \gamma =2$, are the critical exponents characteristic of the universality class to which the fluid belongs. The superscript $\pm$ of the analytical scaling function $f^\pm$ refer to $h_2 > 0$ and $h_2 < 0$, respectively.

The critical phase transition is described by two scaling densities, associated to the fields $h_1$ and $h_2$: the {\it ordering density} (order parameter) $\phi_1$ and the {\it thermal density} $\phi_2$ 
\begin{equation}
 \phi_1 \equiv \left( \dfrac{\partial h_3 }{ \partial h_1}\right)_{h_2} \text{ } , \text{ } \phi_2 \equiv \left( \dfrac{\partial h_3 }{ \partial h_2}\right )_{h_1} .
\end{equation}
The variation of the scaling densities with respect to the scaling fields define the ``susceptibilities'' of the fluid
\begin{eqnarray}
 \chi_1 \equiv \left( \dfrac{\partial \phi_1 }{ \partial h_1}\right)_{h_2} \text{ } , \text{ } \chi_2 \equiv \left (\dfrac{\partial \phi_2 }{ \partial h_2}\right)_{h_1} \text{ } , \nonumber \\   \\ \nonumber \chi_{12} \equiv \left ( \dfrac{\partial \phi_1 }{ \partial h_2}\right)_{h_1} = \left ( \dfrac{\partial \phi_2 }{ \partial h_1}\right)_{h_2}
\end{eqnarray}
respectively known as strong susceptibility, weak susceptibility and cross susceptibility.
For a liquid system, the thermodynamic response functions experimentally accessible are i) the isothermal  compressibility $K_T\equiv -(1/\langle V\rangle)(\partial \langle V\rangle/\partial P)_T$, ii) the isobaric coefficient of thermal expansion $\alpha_P\equiv (1/\langle V\rangle)(\partial \langle V\rangle/\partial T)_P$ and iii) the isobaric specific heat $C_P\equiv (\partial \langle H\rangle/\partial T)_P$. 
In the previous expressions the symbol $\langle ... \rangle$ refers to the thermodynamic average, $V$, $P$, $T$,  and $H$ refer to the volume, pressure, temperature and enthalpy of the system respectively, with $H\equiv E+PV$, where is the $E$ energy.
The quantities $K_T$, $\alpha_P$ and $C_P$ can be expressed as linear combination of the susceptibilities $\chi_1$, $\chi_2$ and $\chi_{12}$. Note that the generalized susceptibilities $\chi_1$ and $\chi_2$, rather than the response functions $K_T$ and $C_P$, are expected to diverge approaching to the critical point with the characteristic exponents $\gamma$ and $\alpha$ respectively.

At the critical point all the scaling fields vanishes
\begin{equation}
 h_1=h_2=h_3 = 0 
\end{equation}
while the coexistence line is given by
\begin{equation}\label{eq_widom}
 h_1=0 .
\end{equation}
The equation (\ref{eq_widom}) defines also a locus that emanates from the critical point into the supercritical region as analytical continuation of the coexistence line. Such locus is, by definition, the Widom line. The line $h_1=0$ corresponds, by construction,  also to the line of maxima of the statistical correlation length of the system $\xi$. Indeed, it corresponds to the locus where the ordering field vanishes, allowing the fluctuations to spread over broader distances.

According to Holten et al. \cite{Holten2012a}, in the vicinity of the liquid-liquid critical point $(T_c,P_c)$, the scaling fields $h_1$ and $h_2$ can be expressed as linear combination of the physical fields $P$ and $T$
\begin{equation}
 h_1 = \Delta \hat{T} + a' \Delta \hat{P}
 \label{h1}
\end{equation}
\begin{equation}
 h_2 = \Delta \hat{P} + b'\Delta \hat{T}
  \label{h2}
\end{equation}
where 
\begin{equation}
 \Delta \hat{T} = \dfrac{T-T_c}{T_c},\qquad \Delta \hat{P} = \dfrac{(P-P_c)V_c}{RT_c}.
\end{equation}
with $V_c$ the critical volume and $R$ the gas universal constant. The coefficient $a'\equiv -d\hat{T}/d\hat{P}$ of Eq. (\ref{h1})  is the slope in the $T$--$P$ plane of the coexistence line (or Widom line) at the critical point, while the coefficient $b'$ is the mixing-coefficient of the {\it mixed-field} order parameter that accounts for the lack of symmetry in the critical density distribution \cite{PhysRevLett.68.193}, \cite{Bianco2014}.
%Bruce and Wilding, Phys. Rev. Lett.
%68, 193 (1992); Bianco and Franzese, Sci. Rep. 4, 4440 (2014)).

%\bibliography{/home/valentino/Dropbox/valentino-giancarlo/Bibliografia_Completa}

\section*{References}
\bibliography{percolation.bib}

\end{document}